\documentclass[letterpaper,twocolumn,10pt]{article}
\usepackage{usenix-2020-09}
\usepackage{booktabs}

\usepackage{tikz}
\usepackage{amsmath}

\usepackage{times,url,color,soul,xspace,enumitem}
\usepackage{graphicx}
\usepackage{caption}
\usepackage{subcaption}
\usepackage{comment}
\usepackage{xspace}
\usepackage[utf8]{inputenc}

\usepackage{cleveref}
\usepackage{hyperref}
\crefformat{section}{\S#2#1#3} 
\crefformat{subsection}{\S#2#1#3}
\crefformat{subsubsection}{\S#2#1#3}

\hypersetup{
    pdftitle={Evolution of Cybersecurity Subdisciplines: A Science of Science Study},
    pdfsubject={Computer Science - Cryptography and Security},
    pdfkeywords={cybersecurity, science of science, gender dynamics, collaboration networks}
}

\begin{document}

\date{}

\title{\Large \bf Evolution of Cybersecurity Subdisciplines: A Science of Science Study}

\author{
{\rm Yao Chen}\\
University of Southampton, UK\\
\texttt{yao.chen@soton.ac.uk}
\and
{\rm Jeff Yan}\\
University of Southampton, UK\\
\texttt{jeff.yan@soton.ac.uk}
}

\maketitle

\begin{abstract} 
The science of science is an emerging field that studies the practice of science itself. We present the first study of the cybersecurity discipline from a science of science perspective. We examine the evolution of two comparable interdisciplinary communities in cybersecurity—the Symposium on Usable Privacy and Security (SOUPS) and Financial Cryptography and Data Security (FC). 
\end{abstract}

\section{Introduction}

The emerging field of "Science of Science" (SciSci)~\cite{fortunato2018science}
is a discipline that places scientific practices under its research lens, aiming to uncover the patterns of scientific activities through quantitative analysis. SciSci integrates research methods and theories from multiple disciplines and employ extensive datasets and toolkits.

SciSci has uncovered some interesting and profound insights~\cite{fortunato2018science}. For example, truly innovative and highly interdisciplinary ideas often 
go unrecognized unless framed within the context of established knowledge~\cite{uzzi2013atypical}.
A scientist's highest-impact work can occur at any point in their career—early, mid, or late—with roughly equal probability~\cite{sinatra2016quantifying}.
Collaborative team research is the norm, and \textit{small} teams tend to be disruptive, but \textit{big} teams have more impact~\cite{milojevic2015, wuchty2007increasing, wu2019large}.
Contrary to funding agencies' claims of support,  interdisciplinary research has consistently lower funding success~\cite{bromham2016interdisciplinary}.

While SciSci strives to identify universal patterns and mechanisms applicable across disciplines, the significant differences between fields highlight the critical importance of discipline-specific research. Variations in cultural norms, practices, data requirements, and skills among disciplines mean that relying solely on universal patterns may not fully capture the nuances of these differences. 

Studying cybersecurity through the SciSci lens is rare. In this paper, we report such a study that examines the evolution of two mainstream cybersecurity subdisciplines from their inceptions, one is the usable security community, represented by the Symposium on Usable Privacy and Security (SOUPS), and the other financial cryptography and security, represented by the conference on Financial Cryptography and Data Security (FC). We have chosen these two communities for the following reasons. First, both are quality venues that have encouraged interdisciplinary work from the very beginning. Second, although both began as niche topics, they have since become mainstream in cybersecurity.

Purely curiosity-driven, we set out to explore the following questions. What can we inform of future research leaders who are going to create and bootstrap a new field? 
What actionable practices and policy can we recommend to the cybersecurity community at large and other stakeholders?

\section{Datasets}

We considered only the main conference papers from the inception year of each conference (2005 for SOUPS and 1997 for FC) to 2023, excluding workshop papers, posters, auxiliary meetings, and other similar contributions. We collected the following metadata: paper titles, author lists, and citation data. Paper titles and author lists were extracted from the DBLP database~\cite{dblp} using Python-based web scraping techniques. Citation data for SOUPS and FC papers was retrieved from Google Scholar.

Additionally, we analyzed award-winning papers from SOUPS (2005–2023) by extracting relevant information from the conference website hosted by USENIX~\cite{usenix}. Notably, no corresponding award paper information was available for FC.

\section{Basic statistics}\label{sec:Basic statistics}

\begin{figure}[h]
\centering
\includegraphics[width=\linewidth]{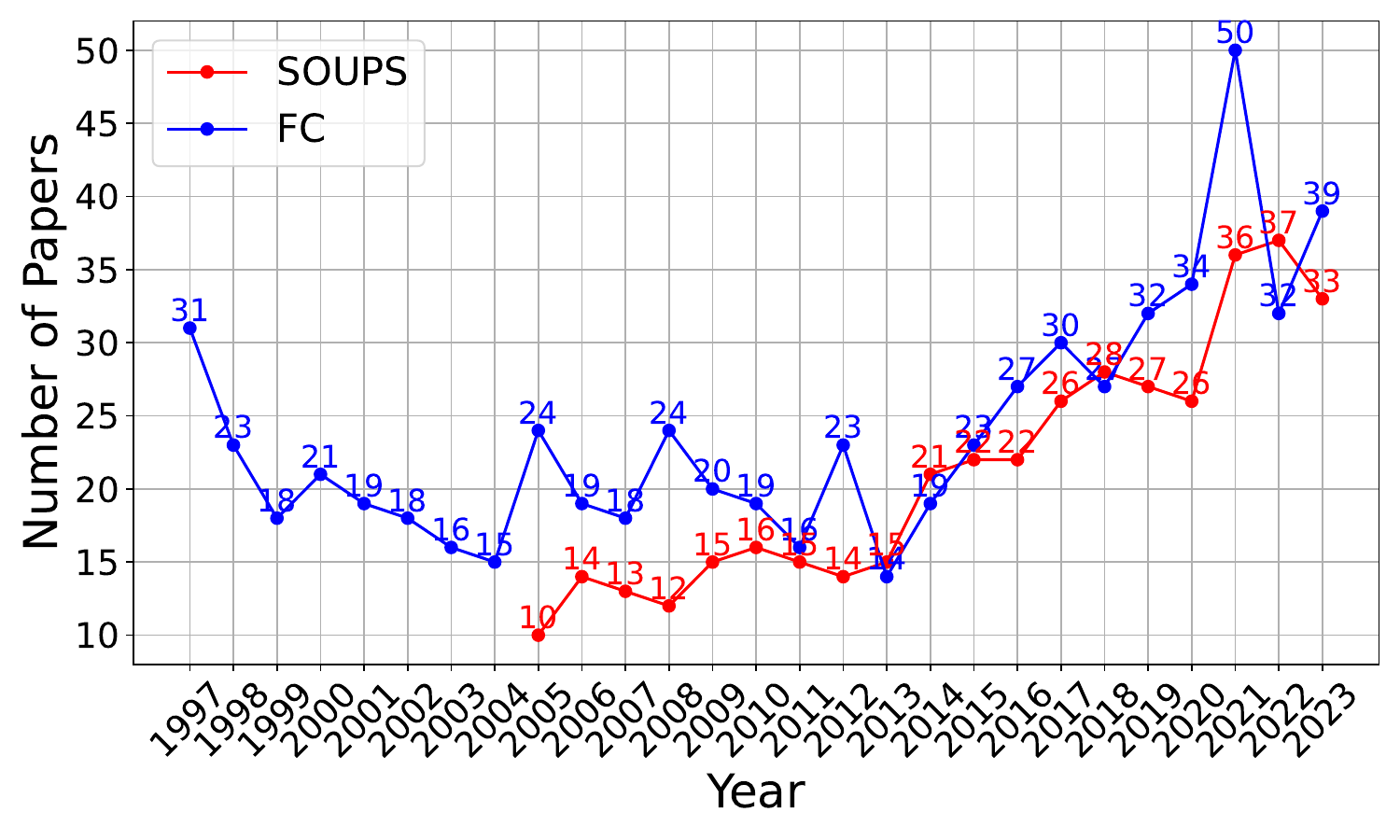}
\caption{The number of papers published each year. The total number of papers: 402 for SOUPS and 651 for FC.}
\label{fig:FC_vs_SOUPS_Publications_Per_Year_with_Larger_Labels}
\end{figure}

\textbf{Publication volume.} 
As Figure \ref{fig:FC_vs_SOUPS_Publications_Per_Year_with_Larger_Labels} suggests, the number of papers published per year has largely followed an upward trend for both SOUPS and FC. For most of the years, FC published more papers than SOUPS each year. In our dataset, the total number of papers is 402 for SOUPS and 651 for FC, respectively.

\textbf{Influence:} The most cited SOUPS paper has a citation count of 1,507. Among the top 10\% most-cited SOUPS papers, their average citation count is 439.9, and the median citation count is 373.5. The top five most-cited SOUPS papers are as follow:
\begin{itemize}
    \item Android Permissions: User Attention, Comprehension, and Behavior (1,507 citations).

   \item The Battle Against Phishing: Dynamic Security Skins (803 citations).

   \item Anti-Phishing Phil: The Design and Evaluation of a Game That Teaches People Not to Fall for Phish (786 citations).
   
   \item "I Regretted the Minute I Pressed Share": A Qualitative Study of Regrets on Facebook (746 citations).

    \item A "Nutrition Label" for Privacy (620 citations).
\end{itemize}

The most cited FC paper has a citation count of 3,012. The top 10\% most-cited FC papers have a mean citation count of 429.26 and a median citation count of 300.
The top five most-cited FC papers are as follow: 

\begin{itemize}
\item Majority Is Not Enough: Bitcoin Mining Is Vulnerable (3,012 citations).

\item Quantitative Analysis of the Full Bitcoin Transaction Graph (1,454 citations).

\item Secure High-Rate Transaction Processing in Bitcoin (1,017 citations).

\item Bitter to Better - How to Make Bitcoin a Better Currency (1,001 citations). 

\item Evaluating User Privacy in Bitcoin (998 citations).
\end{itemize}

The figures clearly demonstrates the quality of the conferences, and that these papers have had an influence far beyond the conferences themselves.

\begin{figure}[h]
\centering
\includegraphics[width=\linewidth]{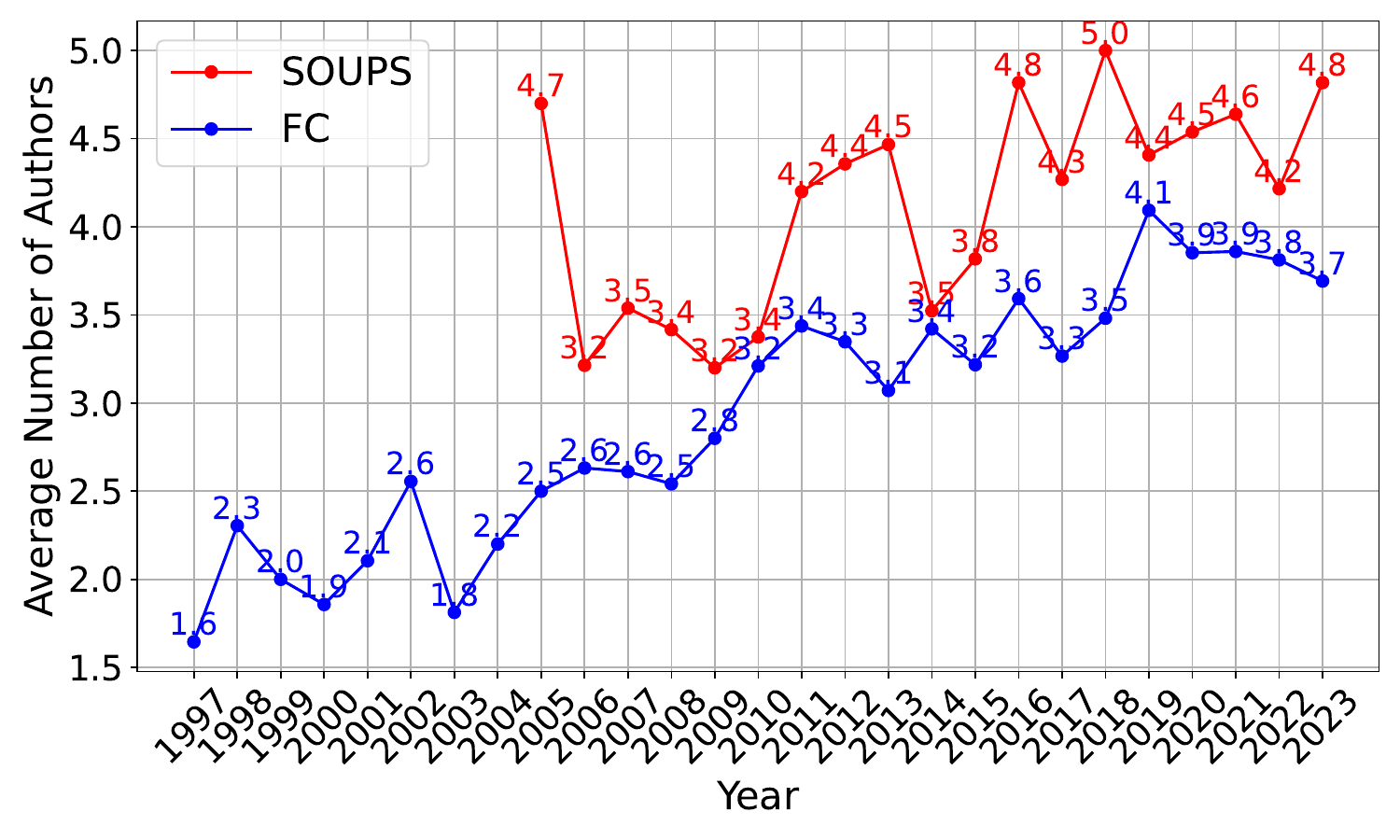}
\caption{The Evolution of Average Team Sizes.).
}
\label{fig:SOUPS_vs_FC_Average_Authors_per_Year}
\end{figure}

\textbf{Team science:} Figure \ref{fig:SOUPS_vs_FC_Average_Authors_per_Year} plots the average number of authors per paper each year for SOUPS and FC. Overall, the average team size followed an upward trend for both conferences. We also calculated the median number of authors per paper each year, and it followed a similar pattern. 
However, SOUPS consistently has a larger average team size than FC. When considering all papers, the average team size is 4.24 for SOUPS and 3.05 for FC, with medians of 4 and 3, respectively.

\begin{figure}[h]
\centering
\includegraphics[width=\linewidth]{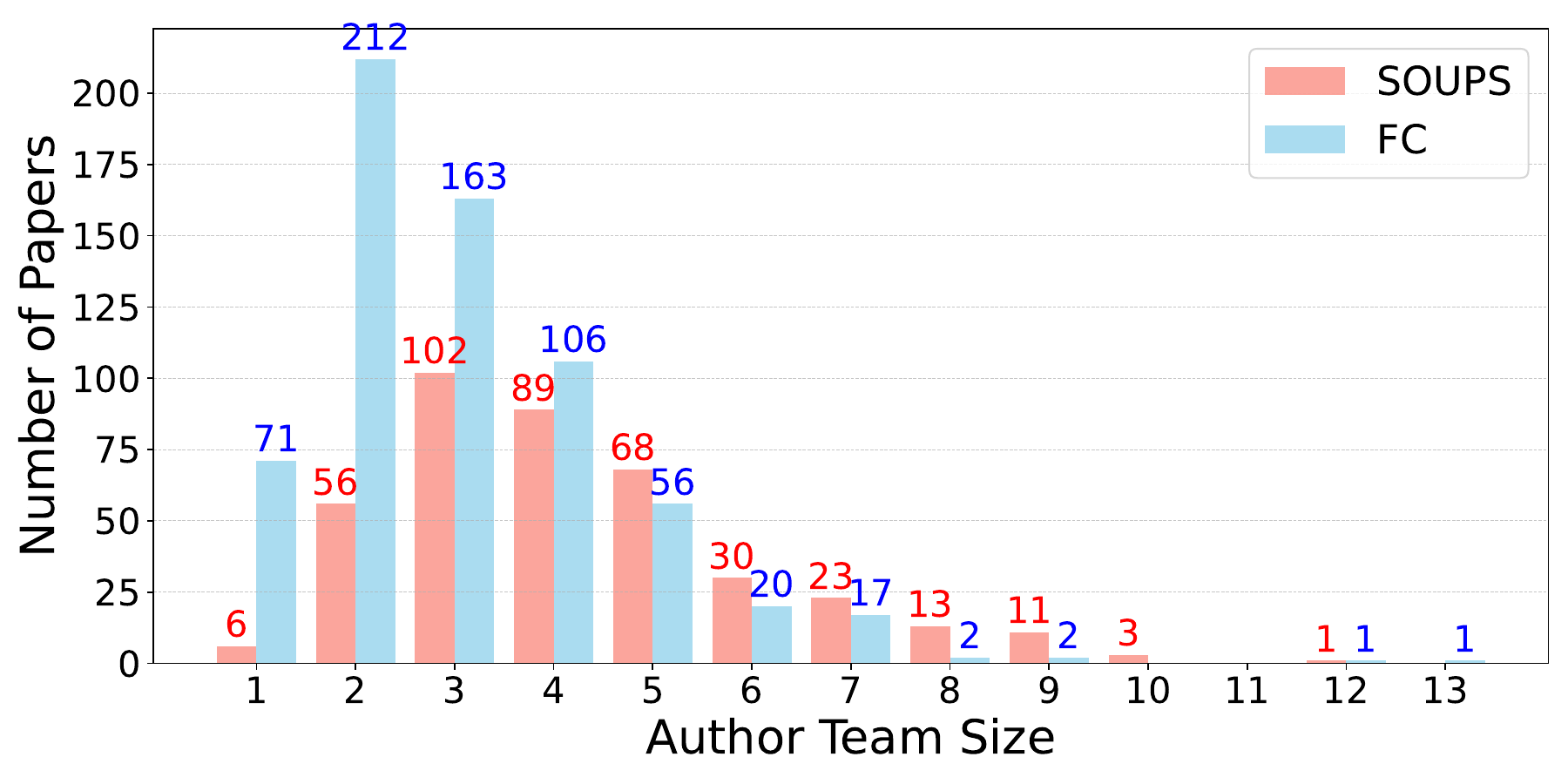}
\caption{Distribution of Papers by Author Team Size. }
\label{fig:Comparison_of_Papers_Distribution_with_Full_XTicks}
\end{figure}

Figure \ref{fig:Comparison_of_Papers_Distribution_with_Full_XTicks} compares the number of papers published by teams of different sizes in SOUPS and FC. We observed two patterns. First, teams with three or four authors are the most common in SOUPS, accounting for 25.37\% and 22.14\% respectively. In contrast, teams of two or three authors are the most common in FC, accounting for 32.57\% and 25.04\% respectively. Second, the number of teams larger than six authors is small, in particular for FC.

\begin{figure*}[h]
\centering
\includegraphics[width=0.9\linewidth]{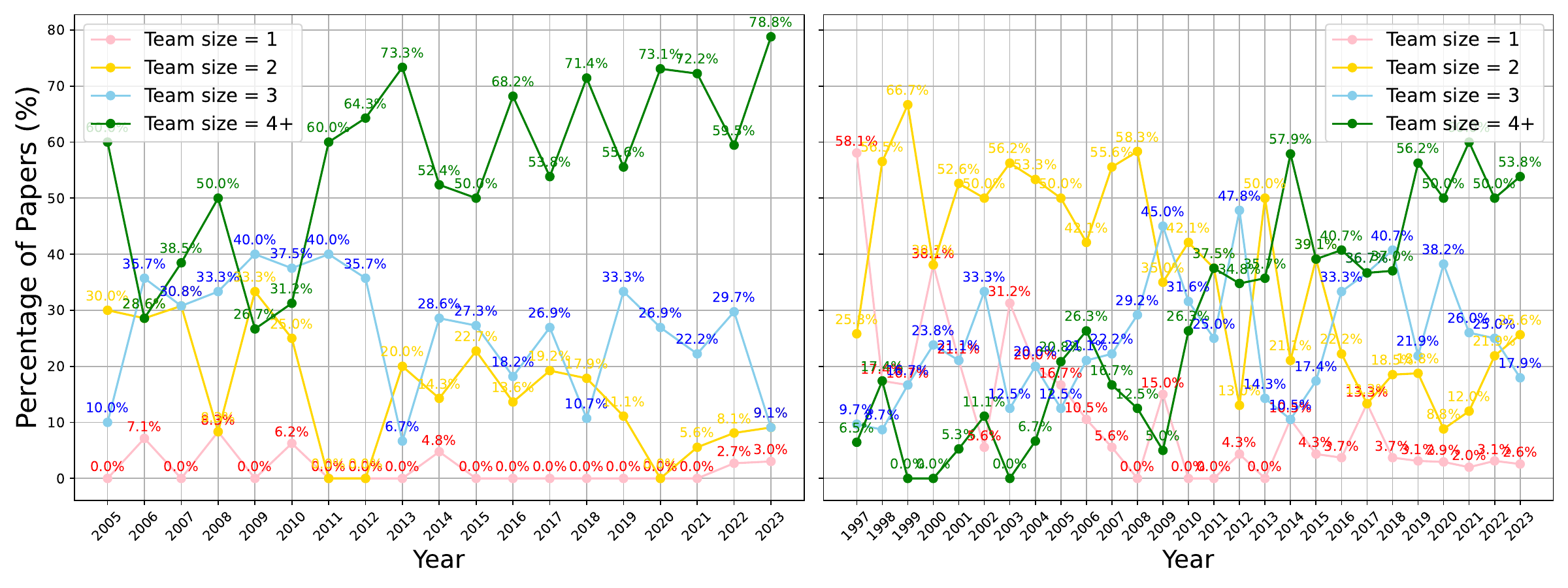}
\caption{Portions of papers produced by teams of different sizes (Left: SOUPS; Right: FC).}
\label{fig:SOUPS_and_FC_Team_Size_Percentage_Distribution}
\end{figure*}

To facilitate the analysis of team size trends, we categorized author teams into four groups: single-author teams, two-author teams, three-author teams, and teams with four or more authors. 
This categorization ensures clarity and simplifies comparative analysis, as teams with more than four authors represent a relatively small proportion of the data. 
Figure \ref{fig:SOUPS_and_FC_Team_Size_Percentage_Distribution} shows the distribution of papers produced by these four team categories each year. 
Overall, the proportion of teams with four or more authors has shown an upward trend in both conferences, reflecting a shared shift toward larger team collaborations. The key difference lies in the timing and magnitude of this trend: SOUPS adopted this pattern earlier and more prominently, reaching 78.8\% in 2023. In contrast, FC was dominated by two-author teams from 1998 to 2008. It is only in recent years that teams with four or more authors have become the most common in FC.

\section{Gender dynamics}\label{sec:Gender Dynamics}
Analyzing gender dynamics in academic research is crucial for understanding gender equality in academia and its development trends, as well as the relationship between gender and high-impact research. Name-based gender classification is an efficient method that provides essential data for studying gender roles in academic publishing, collaboration, and career development. Among these methods, the CCT classifier~\cite{van2023open}, introduced in 2023, stands out for its open-source design based on ensemble learning and approaches the theoretical upper limit of classification accuracy.
However, the original study did not directly evaluate the classifier on a ground truth dataset, but tested its performance indirectly by comparing it to state-of-the-art commercial services, leaving its reliability inadequately validated. 

To address this, we used the actual gender data of SOUPS authors as a ground truth dataset to evaluate the CCT classifier's accuracy and proposed an improvement to enhance prediction performance. The study validated the effectiveness of the improvement and provided practical recommendations for its application. Additionally, we applied the CCT classifier to predict the gender of FC authors.
Finally, we analyzed the gender distribution of authors in SOUPS and FC, examined annual trends, and further explored the evolution of team gender composition and the gender of first authors.

\subsection{Predicting gender by name}\label{sec:Predicting gender by name}
To enable researchers to apply the CCT classifier in gender-related studies, it is important to delve into its theoretical foundations, implementation details, improvement, and performance evaluation. 
It enhances the transparency of the CCT classifier while facilitating its broader adoption in academic and practical applications.

\subsubsection{CCT classifier}
\textbf{Theoretical Foundations. } The CCT classifier, based on Cultural Consensus Theory (CCT),  which was initially developed as an analytical method in anthropology to infer collective consensus from multiple data sources. The classifier estimates the probability (\(z_m\)) of a name (\(m\)) being female while evaluating the reliability or competence (\(c_n\)) of each data source (\(n\)). Unlike simple averaging, it dynamically weights each source's input based on its competence, determined by agreement with the consensus.

CCT's mathematical foundation is Bayes' theorem. 
Given the responses of the sources (\(\mathbf{x}\)) and their competences (\(\mathbf{c}\)), the probability of a name (\(m\)) being gendered female is 
\begin{equation}
    P(y_m = 1 \mid \mathbf{x}, \mathbf{c}) \propto \prod_{n=1}^N P(x_{n,m} \mid y_m = 1, c_n),\label{eq:ym}
\end{equation}
where \(y_m\) is the consensus estimate of the gender associated with 
$m$, where $y_m = 1$ indicates female, and $y_m = 0$ indicates male.
\(x_{n,m}\) is the gender prediction for the name \(m\) by source \(n\); 1 being for female, 0 for male.
 \(P(x_{n,m} \mid y_m = 1, c_n)\) is the likelihood of observing \(x_{n,m}\), given the consensus \(y_m = 1\) and competence \(c_n\). 
By explicitly defining competence \(c_n\) to be the probability that source \(n\) correctly reports the consensus, independent of the name, we can obtain:
\begin{equation}
P(x_{n,m} \mid y_m = 1, c_n) = x_{n,m}c_n + (1 - x_{n,m})(1 - c_n).\label{eq:xnm}
\end{equation}
Substituting Eq.~(\ref{eq:xnm}) into Eq.~(\ref{eq:ym}) and using the notation \(z_m \equiv P(y_m = 1 \mid \mathbf{x}, \mathbf{c})\), the calculation of \(z_m\) is as follows:
\begin{equation}
\begin{aligned}
z_m = 
\prod_{n=1}^N \left[x_{n,m} c_n + (1 - x_{n,m})(1 - c_n)\right] / \\
\bigg(
\prod_{n=1}^N \left[x_{n,m} c_n + (1 - x_{n,m})(1 - c_n)\right] 
+ \\ 
\prod_{n=1}^N \left[x_{n,m}(1 - c_n) + (1 - x_{n,m})c_n\right]
\bigg)
\end{aligned}\label{eq:zm}
\end{equation}

\(c_n\) is determined by its average agreement with the consensus across all queried names and is calculated as:
\begin{equation}
c_n = \frac{1}{M} \sum_{m=1}^M \left[x_{n,m}z_m + (1 - x_{n,m})(1 - z_m)\right],\label{eq:cn}
\end{equation}
where \(M\) is the total number of queried names.

The CCT classifier uses the Expectation Maximization (EM) algorithm to iteratively optimize \(z_m\) and \(c_n\). During the expectation step, the consensus probabilities (\(z_m\)) are updated using the current competence values. During the maximization step, the competences (\(c_n\)) of the data sources are recalculated based on the updated consensus. This process is repeated until convergence, ensuring robust estimates of gender probabilities and data source reliability.

\textbf{Technical Details.} 
The CCT classifier includes a dataset of name-gender associations constructed by integrating 36 public resources. 
A \(p(\textit{gf})\) value, corresponding to \(z_m\) in Eq.~(\ref{eq:zm}), is calculates with the EM algorithm  for each name, indicating the probability of the name being gendered female.

The CCT classifier supports two methods for gender predication. The first method uses the \(p(\textit{gf})\) value only:
a name is predicated as female if and only if its \(p(\textit{gf}) \geq 0.5\).

The second method, i.e. its default method, is more sophisticated. A gender uncertainty (\(u\)) is introduced to characterise if a name has clear gender characteristics as follows:

\begin{equation}
u = 0.5 - \lvert 0.5 - p(\textit{gf}) \rvert \label{eq:uncertainty}
\end{equation}

Then, a predication is determined as follows. 
     \begin{itemize}
    \item If \(u \leq \tau\), a predefined uncertainty threshold, the name is classified based on \(p(\textit{gf})\): if \(p(\textit{gf}) \geq 0.5\), the name is classified as gendered female; otherwise,  
    gendered male.
    \item If \(u > \tau\) or the name is not found in the dataset, the predication outputs \textit{unknown}. 
\end{itemize}

\subsubsection{Improvement}

Before searching the name-gender association dataset, the CCT classifier preprocesses names by removing diacritical marks, converting them to lowercase, and trimming whitespace. However, these steps do not accommodate specific name formats, such as those with initials or abbreviations, which may lead to classification errors. Therefore, an improvement was introduced to handle names beginning with a single letter or a letter followed by a period (e.g., “J.” or “A.”). This step removes the first word of such names, retaining the part relevant to gender prediction. For example, “J. Doug Tygar” was transformed into “Doug Tygar”.

This improvement effectively eliminates the interference caused by initials, which often lack gender information, allowing the classifier to focus more precisely on the gender-relevant components of names. Additionally, it enhances the name matching success rate of the CCT classifier. However, while this improvement mitigates the influence of noise on classification results, it cannot completely eliminate non-gender-related components in names.

\subsubsection{Evaluation}\label{Algorithm Evaluation}

\begin{table}[h!]
    \centering
    \renewcommand{\arraystretch}{1.3} 
    \small
    \begin{tabular}{lcccc}
        \toprule
        \textbf{$\tau$} & \textbf{Improvement} & \textbf{\boldmath\( \text{Count}_{\text{unclassified}} \)} & \textbf{\boldmath\( \text{Acc}\)} &  \textbf{\boldmath\( \text{Acc}' \)}  \\
        \midrule
        0.1 & No & 187 & 81.10\%  & 98.30\% \\
            & Yes & \textcolor{blue}{184} &  \textcolor{blue}{82.69\%}  & \textcolor{blue}{98.74\%} \\
        0.2 & No & 132 &  85.59\%  & 97.65\% \\
            & Yes & \textcolor{blue}{126} &  \textcolor{blue}{87.31\%}  & \textcolor{blue}{98.06\%} \\
        0.3 & No & 106 &  87.56\%  & 97.20\% \\
            & Yes & \textcolor{blue}{101} &  \textcolor{blue}{89.23\%} & \textcolor{blue}{97.68\%} \\
        \bottomrule
    \end{tabular}
    \caption{Evaluating the CCT default method with SOUPS ground-truth data. }
    \label{tab:algorithm_evaluation}
\end{table}

\textbf{Accuracy.} We evaluated the CCT classifier using three metrics:

\begin{itemize}
    \item The number of names that the classifier fails to predicate gender, denoted by $c_u$. 

    \item \textbf{Classified Accuracy (\( \text{Acc}' \))}: The percentage of correctly predicted names ($p$) among those that have a non-unknown classification, denoted by $c$.

    \item \textbf{Overall Accuracy (\( \text{Acc} \))}: The percentage of correctly classified names (\(p\)) out of the total names. Acc = $p/(c_u + c)$.

\end{itemize}

Table~\ref{tab:algorithm_evaluation} compares the default method's performance under different uncertainty thresholds (\(\tau\)), with and without our improvement. For \(\tau = 0.1\), the improvement reduces 
$c_u$ from 187 to 184, increases \( \text{Acc} \) from 81.10\% to 82.69\%, and improves \( \text{Acc}' \) from 98.30\% to 98.74\%, demonstrating enhanced classification performance.
As \(\tau\) increases, 
$c_u$ decreases, and \( \text{Acc} \) improves, as higher \(\tau\) values reduce the classifier’s strictness, allowing more names to be classified as male or female. However, this also increases the likelihood of misclassifications due to the inclusion of names with higher uncertainty, slightly lowering \( \text{Acc}' \).

We also evaluated the $p(gf)$ method's performance.
With our improvement,  \( \text{Acc}\) increased from 91.49\% to 93.27\%, 
decreased from 39 to 27, and \( \text{Acc}' \) improved from 94.95\% to 95.57\%.

\textbf{Coverage} 
is the ratio of classifiable names ($c$) out of the total number of names. Value $c$ is directly influenced by \(\tau\), as \(\tau\) determines the strictness of the classification standard. A higher \(\tau\) value indicates a more lenient classification standard, allowing more names to be classified but potentially reducing classification accuracy.

\begin{table*}[h!]
    \centering
    \renewcommand{\arraystretch}{1.3} 
    \small
    \begin{tabular}{lccccccccccccccc}
        \toprule
        \textbf{$\tau$} & 0.3 & 0.28 & 0.26 & 0.24 & 0.22 & 0.2 & 0.18 & 0.16 & 0.14 & 0.12 & 0.1 & 0.08 & 0.06 & 0.04 & 0.02 \\
        \midrule
        \textbf{\textit{Coverage}\_1 (\%)} & 90 & \textbf{89} & 89 & \textbf{88} & 88 & \textbf{87} & \textbf{86} & 85 & 84 & 84 & 82 & \textbf{80} & 79 & 76 & 69 \\
        \textbf{\textit{Coverage}\_2 (\%)} & 90 & \textbf{90} & 89 & \textbf{89} & 88 & \textbf{88} & \textbf{87} & 85 & 84 & 84 & 82 & \textbf{81} & 79 & 76 & 69 \\
        \bottomrule
    \end{tabular}
    \caption{Gender uncertainty threshold ($\tau$) vs coverage, where \textit{Coverage}$_1$ and \textit{Coverage}$_2$ represent the results before and after applying the proposed improvement, respectively. The bold values indicate differences between \textit{Coverage}$_1$ and \textit{Coverage}$_2$.}
    \label{tab:soups_CCT_Classifier_Thresholds}
\end{table*}
To automatically determine a suitable \(\tau\) value, the CCT classifier introduces a "tune" function, which uses a heuristic approach to find the smallest \(\tau\) ensuring \textit{Coverage} exceeds 85\%. 
Table \ref{tab:soups_CCT_Classifier_Thresholds}  shows that the improvement slightly increases \textit{Coverage} across most thresholds. For example, at \(\tau = 0.08\), \textit{Coverage} improves from 80\% (\textit{Coverage}$_1$) to 81\% (\textit{Coverage}$_2$). This improvement is achieved without compromising classification accuracy, as it enhances the name-matching success rate in the CCT classifier.

For high \textit{Coverage} scenarios, such as 90\%, the proposed improvement shows clear advantages. Without it, a \(\tau\) value of 0.3 is needed to achieve 90\% \textit{Coverage}, while with the improvement, the same \textit{Coverage} is achieved at \(\tau = 0.28\). Lower \(\tau\) values mean stricter criteria, reducing misclassification risk. Thus, the improvement allows the CCT classifier to maintain the same \textit{Coverage} with a lower \(\tau\), minimizing misclassification.

\subsubsection{Recommendations}

If the goal is to maximize the number of classified names
and maintain a good predication accuracy, we recommended to use the predication method that is based only on the \(p(\textit{gf})\) values. 
If the goal is to prioritize a very high accuracy of gender predication but sacrifice the coverage to an extent, we recommend using the CCT's default prediction method.

\subsection{Gender determination for SOUPS and FC}

To ensure reliable gender analysis for SOUPS authors, we directly used their actual gender data. Based on prior evaluations, the CCT classifier demonstrated excellent performance in gender classification. Therefore, we applied it to the FC author dataset to automatically identify genders, significantly reducing manual verification workload. Using the "tune" function, we set a suitable threshold of \( \tau = 0.14 \), achieving \textit{Coverage} of 85\%, which balanced high \textit{Coverage} with strict classification standards.

To further improve accuracy, we integrated the actual gender data of SOUPS authors into the FC dataset. The analysis revealed 74 SOUPS authors in the FC dataset, with 10 names unclassified. For the remaining 64 names classified as male or female, predictions perfectly matched actual genders.
These results highlight the effectiveness of data integration, reducing the number of unknown genders in the FC dataset, and confirm the reliability of the CCT classifier, which achieved 100\% accuracy for classified names. For names the classifier failed to classify, we manually verified their genders to ensure analysis reliability.

\subsection{Gender of authors}

\begin{figure}[h]
  \centering
  \includegraphics[width=\linewidth]{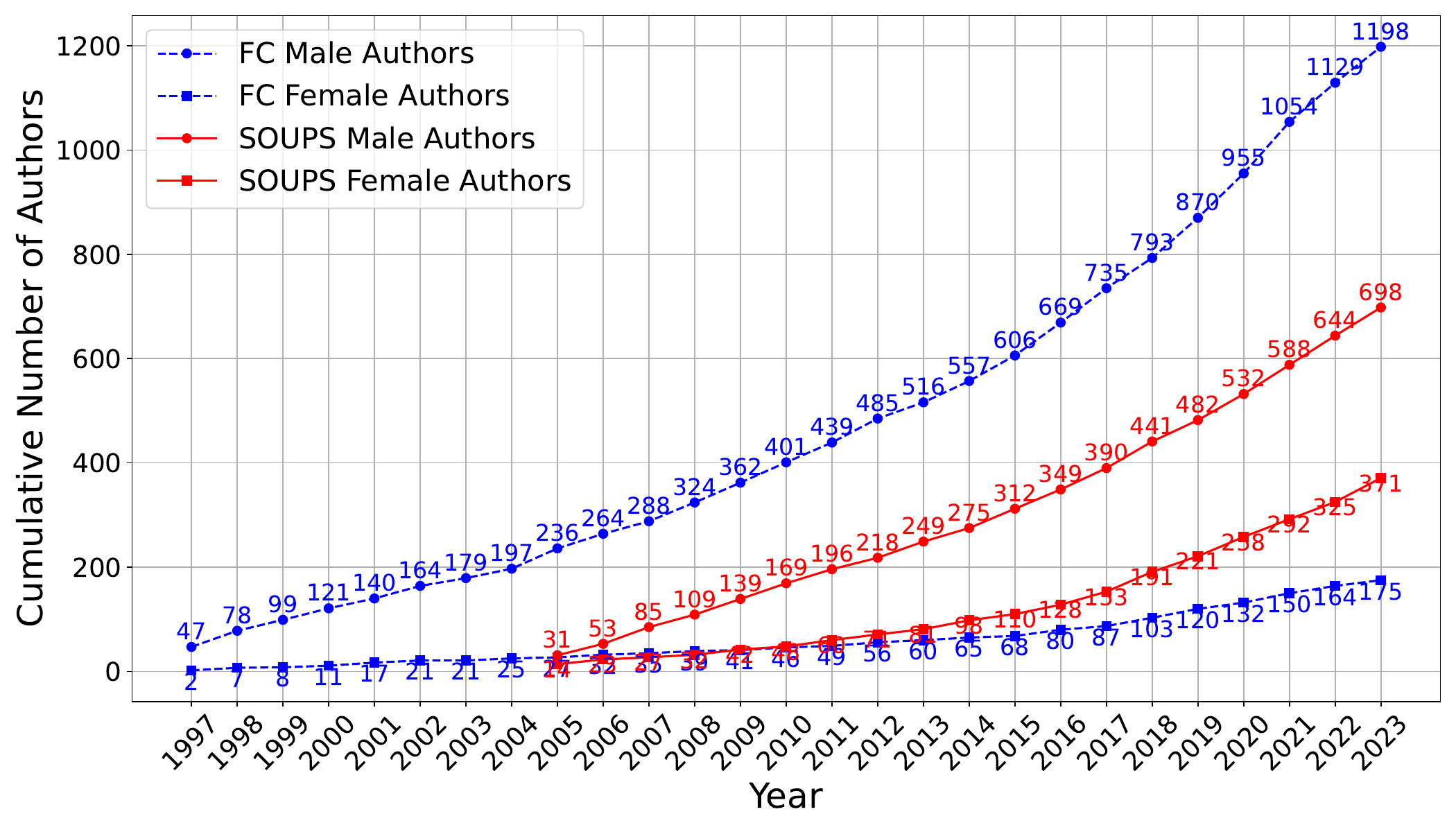}
  \caption{
      Cumulative counts of male and female authors. 
      }
\label{fig:Cumulative_Number_of_Male_and_Female_Authors_FC_vs_SOUPS}
\end{figure}

To explore the gender distribution of authors in SOUPS and FC, we analyzed the cumulative numbers and annual proportions of male and female authors in both conferences. In our dataset, SOUPS had a total of 1069 authors, with male authors accounting for 65.3\% and female authors for 34.7\%. In contrast, FC had 1373 authors, with male authors comprising 87.25\% and female authors only 12.75\%. These figures indicate that SOUPS is more gender balanced than FC.

Figure \ref{fig:Cumulative_Number_of_Male_and_Female_Authors_FC_vs_SOUPS} illustrates the cumulative growth of male and female authors in both conferences over time. 
The comparison shows that the number of male authors in FC and SOUPS has consistently been higher than that of female authors, with this gender gap being particularly pronounced in FC.
Additionally, the steeper slope of the female authors' curve in SOUPS indicates that the growth rate of female authors in SOUPS is higher than in FC.

\begin{figure}[h]
\centering
\includegraphics[width=\linewidth]{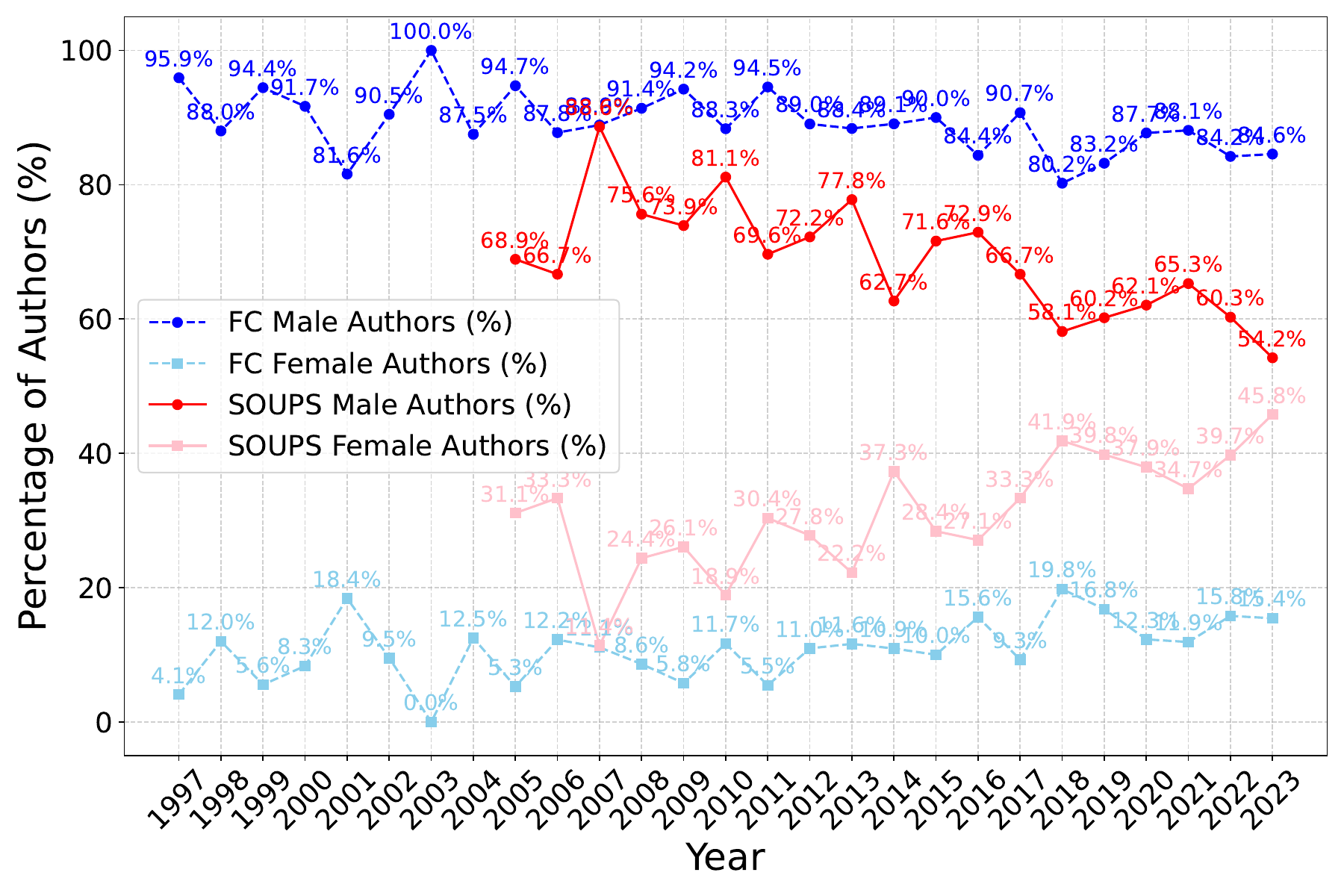}
\caption{Proportions of male and female authors. 
}
\label{fig:Male_and_Female_Authors_FC_vs_SOUPS_Percentage_with_Labels}
\end{figure}

Figure \ref{fig:Male_and_Female_Authors_FC_vs_SOUPS_Percentage_with_Labels} shows the proportion of male and female authors every year. 
Unlike the cumulative trends shown in Figure \ref{fig:Cumulative_Number_of_Male_and_Female_Authors_FC_vs_SOUPS}, Figure \ref{fig:Male_and_Female_Authors_FC_vs_SOUPS_Percentage_with_Labels} provides a year-by-year perspective, offering a clearer view of the dynamic changes in author gender diversity at both conferences.
Overall, the proportion of female authors has increased at both conferences. However, this trend is more pronounced in SOUPS, with female authors accounting for 45.8\% in 2023. In contrast,  the growth of female authors in FC has been limited, with the proportion of female authors consistently remaining below 20\%.

\subsection{Gender composition of teams}

As shown in Figure \ref{fig:Comparison_of_Papers_Distribution_with_Full_XTicks}, the number of papers with teams larger than five authors is relatively small. 
Besides, the median team sizes for SOUPS and FC are 4 and 3, respectively.
Therefore, we define two broad categories of team sizes: \textit{small} teams (three or fewer authors) and \textit{big} teams (more than three authors), to capture the general trends in collaboration scale.

\begin{figure*}[h]
\centering
\includegraphics[width=0.9\linewidth]{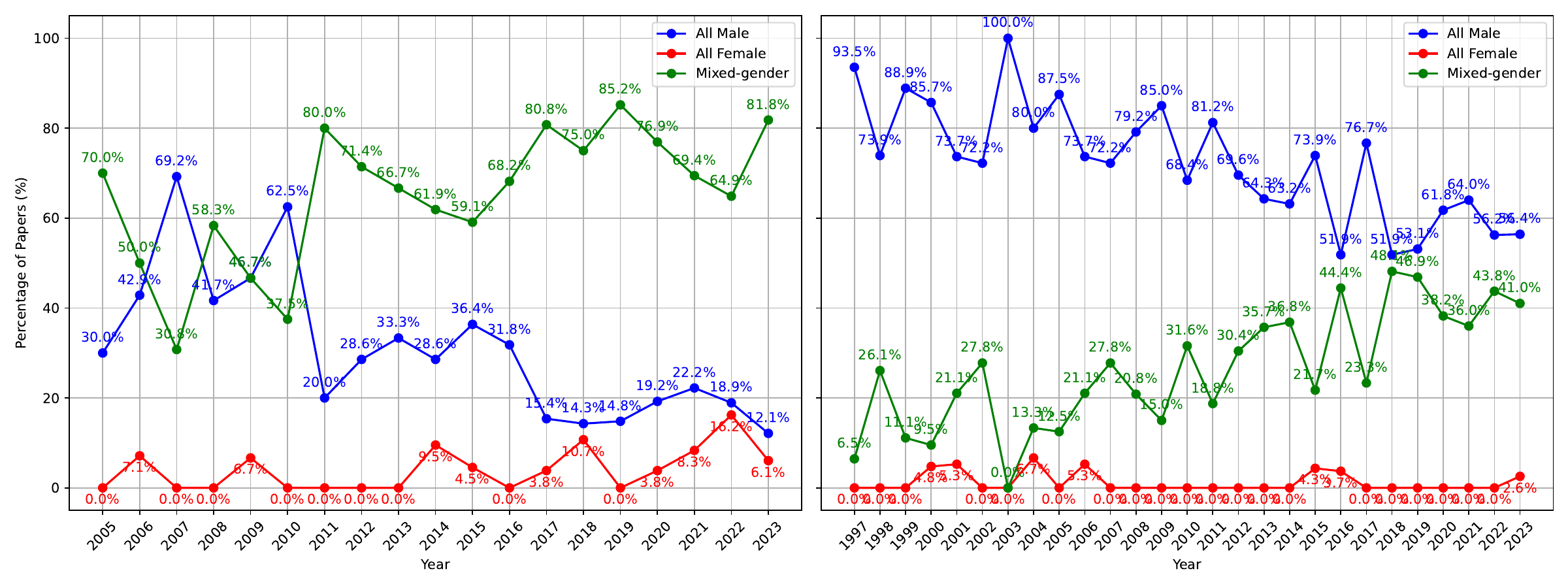}
\caption{Evolution of team gender composition (Left: SOUPS; Right: FC).}
\label{fig:Author_Composition_by_Gender_Percentage_SOUPS_and_FC}
\end{figure*}

To analyze the evolution of gender composition in teams, 
we categorizes teams into three types: all-male teams, all-female teams, and mixed-gender teams.
In our dataset, the counts for these three categories in SOUPS are 109, 21, and 272, respectively, while the corresponding values for FC are 460, 7, and 184. 
Based on these figures, we can know that mixed-gender teams are the most common in SOUPS, accounting for 67.66\%, while  the most common gender composition in FC teams is all male teams, accounting for 70.67\%.
Additionally, all-female teams are relatively rare in both conferences, accounting for 5.22\% in SOUPS and 1.08\% in FC.

Figure \ref{fig:Author_Composition_by_Gender_Percentage_SOUPS_and_FC} illustrates the distribution of team gender composition in SOUPS and FC. Both conferences exhibit similar overall trends: the proportion of mixed-gender teams shows a general increase, while the proportion of all-male teams demonstrates a general decline. 
Analyzing the evolution of the most common team gender composition in these two conferences reveals that, in SOUPS, the proportions of all-male and mixed-gender teams were relatively close in the early years. Starting from 2011, the proportion of mixed-gender teams became significantly higher than the other two types. In contrast, all-male teams have consistently been the most common gender composition in FC. However, with the continuous rise in the proportion of mixed-gender teams, their proportion in FC is approaching that of all-male teams.

\begin{figure*}[h]
\centering
\includegraphics[width=0.9\linewidth]{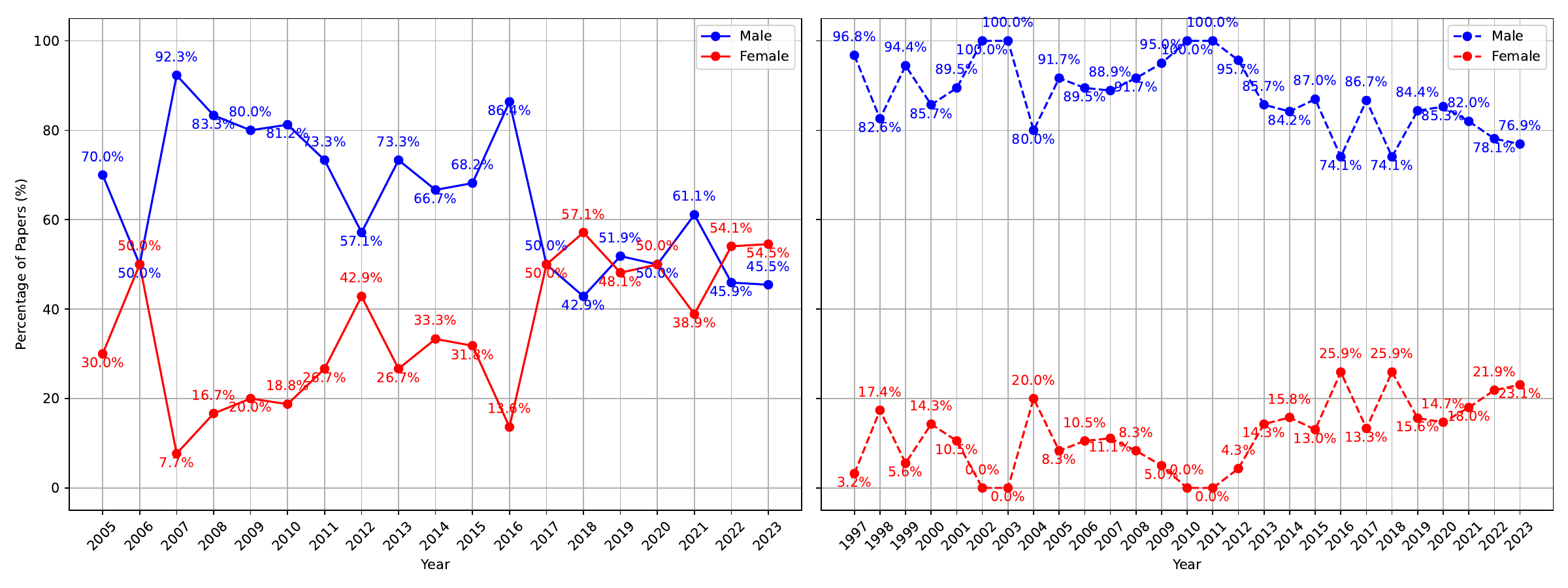}
\caption{Evolution of first author genders (Left: SOUPS; Right: FC).}
\label{fig:Percentage_of_Papers_by_First_Author_Gender_SOUPS_and_FC}
\end{figure*}

To clearly illustrate the gender distribution of the first author in the team over time, Figure \ref{fig:Percentage_of_Papers_by_First_Author_Gender_SOUPS_and_FC} presents the annual proportions of papers authored by male and female first authors. Overall, the proportion of female first authors has shown an increasing trend in both conferences.
Analyzing the gender gap in first authorship for the two conferences, in SOUPS, the proportion of male first authors consistently exceeded that of female first authors before 2017. However, starting in 2017, the proportions of male and female first authors began to converge. Notably, in 2018, 2022, and 2023, the proportion of female first authors surpassed that of male first authors.
In contrast, the gender gap for first authors in FC has remained significant. Although the proportion of female first authors in FC has been increasing, it has never exceeded 26\%.

\section{Research Impact and Excellence}\label{sec:Key Indicators of Research Impact and Excellence}

The three dimensions of influence, productivity, and quality provide a comprehensive framework for evaluating research performance: 
influence reflects the scope of dissemination of research outcomes; productivity represents the level of activity among community members; quality emphasizes the academic standards of the research outcomes. By analyzing these indicators, we can gain insights into the research dynamics of SOUPS and FC while also providing valuable information to researchers to support decision-making in team building.

\subsection{Influence}\label{sec:Influence}

This section uses the citation counts of papers as a metric to evaluate their influence, analyzing all published papers  to identify patterns between team characteristics and research impact, and to uncover significant differences between the SOUPS and FC communities.
Additionally, the top 10\% most-cited papers are examined to explore the characteristics of high-impact research.
The analysis focuses on three dimensions: (1) team size, (2) team gender composition, and (3) the gender of the first author.

\textbf{Comparison of the influence between SOUPS and FC: } According to our dataset, the average citation count for SOUPS is 95.94, compared to 77.70 for FC, indicating that SOUPS has a greater research impact.

\textbf{Team Size:} 
Since teams with six or more authors are rare, this analysis focuses on teams with one to five authors. 
Table \ref{tab:soups_fc_teamsize_citation} presents the proportion of papers and the average citation counts for five different team sizes.
Table \ref{tab:soups_fc_teamsize_citation} highlights three interesting observations:
First, in SOUPS, the most common team sizes are three-author and four-author teams, which performed relatively well, with average citation counts of 90.78 and 91.27, respectively. However, in FC, the most common team sizes are two-author and three-author teams, which achieved the highest citation counts, at 93.87 and 77.77.
Second, although single-author teams are uncommon in both SOUPS and FC, their citation performance differs significantly. In SOUPS, single-author teams performed the best, with an average citation count of 156.83, while in FC, they performed the worst, with only 43.65 citations.
Finally, two-author teams showed strong citation performance in both SOUPS and FC. In SOUPS, they ranked second, while in FC, they not only ranked first but were also the most common team size.

\begin{table}[htbp]
\centering
\renewcommand{\arraystretch}{1.3} 
\setlength{\tabcolsep}{2pt}
{\normalsize 
\begin{tabular}{lccccc}
\toprule
\textbf{Team size} &
1 & 2 & 3 & 4 & 5 \\ 
\midrule
\textbf{\% papers (SOUPS)} & 1.49 & 13.93 & \textcolor{blue}{25.37} & \textcolor{blue}{22.14} & 16.92\\
\textbf{citations} &  \textcolor{red}{156.83} & 106.32 & 90.78 & 91.27 & 71.03 \\ 
\midrule
\textbf{\% papers (FC)} & 10.91  & \textcolor{blue}{32.57} & \textcolor{blue}{25.04} & 16.28 & 8.60  \\ 
\textbf{citation} & \textcolor{red}{43.65} & 93.87  & 77.77 & 59.91 & 77.55  \\ 
\bottomrule
\end{tabular}
}
\caption{Average citation counts vs team sizes. 
}
\label{tab:soups_fc_teamsize_citation}
\end{table}

\begin{table}[t] 
\centering
\renewcommand{\arraystretch}{1.3} 
\setlength{\tabcolsep}{2pt} 
\begin{tabular}{lccccc}
\hline
\textbf{Team Size} & \textbf{Composition} & \multicolumn{2}{c}{\textbf{Percentage}} & \multicolumn{2}{c}{\textbf{Average Citations}} \\
\cmidrule(lr){3-4} \cmidrule(lr){5-6}
 & & \textbf{SOUPS} & \textbf{FC} & \textbf{SOUPS} & \textbf{FC} \\
\hline
Big & Mixed-gender & \textcolor{red}{47.26}  & 13.82 & 98.81 & 82.38 \\  
& All male & 10.70 & 17.67 & 77.00  & 65.16 \\  
& All female & 1.24 & 0 & 65.80  & 0 \\  
Small  & Mixed-gender & 20.40 & 14.44 & \textcolor{red}{103.27}  & \textcolor{red}{91.61} \\ 
& All male & 16.42 & \textcolor{red}{53.0} & 97.74  & 77.74  \\  
& All female & 3.98 & 1.08 & 77.25  & 34.71 \\  
\hline
\end{tabular}
\caption{
Team gender composition, team size and citation counts. The largest value in each column is highlighted in red.
}
\label{tab:Team Proportion_team_citation}
\end{table}

\textbf{Team gender composition: }By calculating the average citation counts for the three types of team gender composition in SOUPS and FC, we found that mixed-gender teams achieved the highest average citations in both conferences (SOUPS: 100.15, FC: 87.09).
To provide a more detailed analysis, Table \ref{tab:Team Proportion_team_citation} provides a classification of teams based on size (\textit{big} team vs \textit{small} team) and gender composition. It presents both the percentage of papers and the average citation counts for each category. 
From Table \ref{tab:Team Proportion_team_citation}, the most common team types differ between the two conferences: in SOUPS, mixed-gender \textit{big} teams are the most common (47.26\%), while in FC, all-male \textit{small} teams are the most common (53\%).
Notably, mixed-gender \textit{small} teams achieve the highest average citation counts in both conferences (SOUPS: 103.27, FC: 91.61).

\begin{table}[t] 
\centering
\renewcommand{\arraystretch}{1.3} 
\setlength{\tabcolsep}{2pt} 

\begin{tabular}{lccccc}
\hline
\textbf{Team Size} & \textbf{1st Author} & \multicolumn{2}{c}{\textbf{Percentage}} & \multicolumn{2}{c}{\textbf{Average Citations}} \\
 \cmidrule(lr){3-4} \cmidrule(lr){5-6}
 & & \textbf{SOUPS} & \textbf{FC} & \textbf{SOUPS} & \textbf{FC} \\
\hline
Big & Male & \textcolor{red}{36.32}  & 25.96 &  94.42 & 72.91 \\  
& Female & 22.89 & 5.53 & 93.78  &  71.81 \\  
Small  & Male & 24.63 &  \textcolor{red}{60.98} & 90.15  & 77.65 \\ 
& Female & 16.17 & 7.53 &  \textcolor{red}{111.23}  & \textcolor{red}{98.90}  \\  
\hline
\end{tabular}
\caption{
First author gender, team size and citation counts. 
The largest value in each column is highlighted in red.
}
\label{tab:first author gender_citation}
\end{table}

\textbf{First author gender: }
According to our dataset, male first authors are more common in both conferences (SOUPS: 60.95\%, FC: 86.94\%), with average citation counts of 92.69 and 76.24, respectively. Female first authors, though less common, have higher average citation counts of 101.01 in SOUPS and 87.42 in FC.
Table \ref{tab:Team Proportion_team_citation} provides a detailed classification of teams based on team size (\textit{big} teams and \textit{small} teams) and first author gender.
The most common team types differ between the two conferences: In SOUPS, male first-author \textit{big} teams are the most common (36.32\%), whereas in FC, male first-author \textit{small} teams dominate (60.98\%).
Notably, in both conferences, female first-author \textit{small} teams exhibit the highest average citation counts (SOUPS: 111.23, FC: 98.90).

\begin{table*}[h!]
    \centering
    \renewcommand{\arraystretch}{1.3} 
    \small
    \setlength{\tabcolsep}{8pt}       
    \begin{tabular}{@{}lcccccc@{}}
        \toprule
        \textbf{
        } & \textbf{Team size} & \multicolumn{3}{c}{\textbf{Team Composition}} & \multicolumn{2}{c}{\textbf{First Author }} \\ \cmidrule(lr){3-5} \cmidrule(lr){6-7}
         & \textbf{
         }  & \textbf{Mixed-Gender} & \textbf{All Male} & \textbf{All Female} & \textbf{Male} & \textbf{Female} \\ 
        \midrule
        \textbf{SOUPS} & Big & \textcolor{red}{47.50\%} & 5.00\% & 0.00\% & \textcolor{red}{32.50\%} & 20.00\% \\ 
         & Small & 25.00\% & 20.00\% & 2.50\% & 22.50\% & 25.00\% \\ 
        \textbf{FC} & Big & 18.46\% & 16.93\% & 0.00\% & 30.77\% & 4.62\% \\ 
         & Small & 16.92\% & \textcolor{red}{47.69\%} & 0.00\% & \textcolor{red}{56.92\%} & 7.69\% \\ 
        \bottomrule
    \end{tabular}
    \caption{Top 10\% most cited papers: team size, gender composition and the first author's gender.
    }    \label{tab:combined_gender_composition}
\end{table*}

\textbf{Top 10\% most-cited papers: } 
We replicate the above analysis on these top-cited papers in terms of team size (small vs big), team gender composition, and the first author gender as in 
Table \ref{tab:combined_gender_composition}. Clearly, 
this analysis derives the same patterns of common team types as we observed on the overall data. These two analyses corroborate the validity of each other.

\subsection{Productivity}

\begin{figure}[h]
\centering
\includegraphics[width=0.9\linewidth]{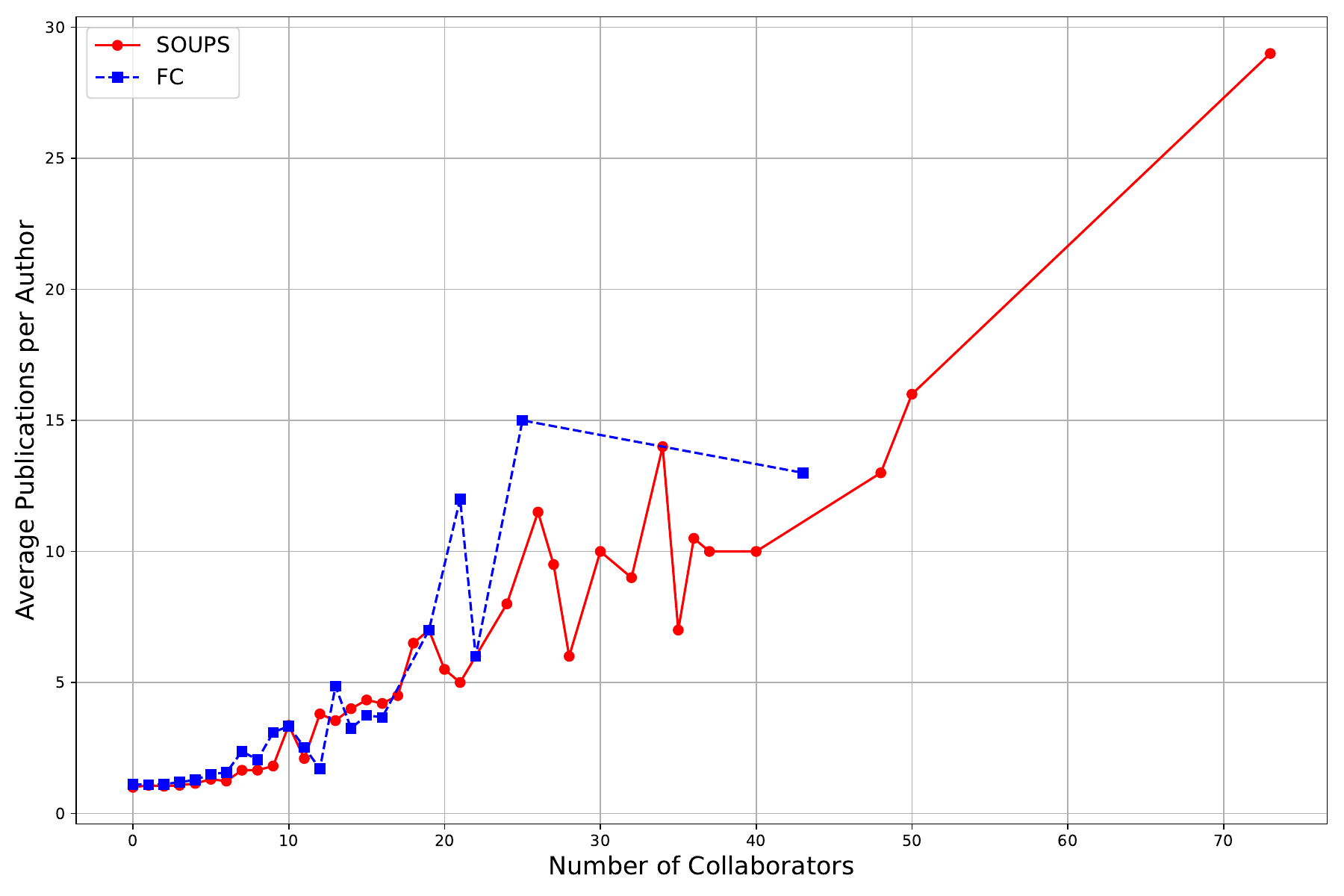}
\caption{The correlation between productivity and the number of collaborators.
}
\label{fig:Comparison of Average Publications by Number of Collaborators (Left: SOUPS; Right: FC)}
\end{figure}

Figure \ref{fig:Comparison of Average Publications by Number of Collaborators (Left: SOUPS; Right: FC)} shows the relationship between the number of collaborators and the average number of publications per author. Figure \ref{fig:Comparison of Average Publications by Number of Collaborators (Left: SOUPS; Right: FC)} indicates an overall positive correlation: authors with more collaborators tend to have higher average academic output (number of publications). We define the author with the highest number of collaborators as the "top collaborator." Figure \ref{fig:Comparison of Average Publications by Number of Collaborators (Left: SOUPS; Right: FC)} reveals a significant difference between the two conferences: in SOUPS, the top collaborator has 73 collaborators, while in FC, this number is 43. This suggests that the SOUPS community has a broader academic collaboration network.

\begin{figure*}[h]
\centering
\includegraphics[width=0.9\linewidth]{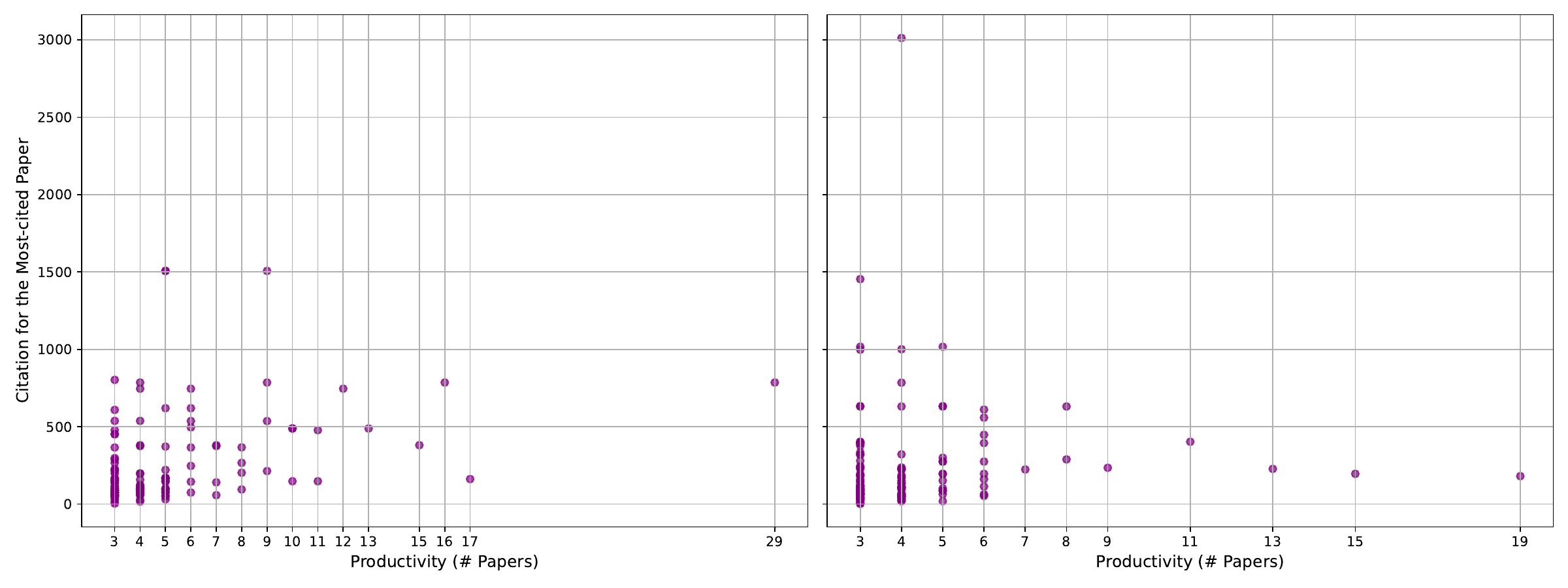}
\caption{The highest citation count of a single paper by the top 10\% most productivity authors (Left: SOUPS; Right: FC). 
High productivity does not necessarily correlate with a high citation count for an author’s most-cited paper. Some authors have many papers, but their papers are not highly cited.
 }
\label{fig:High Productivity Authors and Citations}
\end{figure*}

Figure \ref{fig:High Productivity Authors and Citations} shows the relationship between the number of papers published by the top 10\% most productive authors and the citation count of their most-cited paper. Overall, even authors with the highest number of publications do not necessarily have the highest citation counts for their most-cited paper, indicating no clear positive correlation between productivity and highest citation count.
From Figure \ref{fig:High Productivity Authors and Citations}, it is evident that in SOUPS, the two most-cited authors published only 5 and 9 papers, respectively, while in FC, the most-cited author published just 4 papers. This suggests that authors with fewer publications can still achieve significant citation impact.
Moreover, Figure \ref{fig:High Productivity Authors and Citations} clearly shows that most prolific authors have their highest citation counts concentrated below 500, even with substantial publication output. This trend may reflect the following: prolific authors often distribute their research efforts across multiple directions, making it challenging for any single paper to achieve exceptionally high citation counts. In contrast, authors with lower productivity are more likely to focus on a small number of high-quality papers, resulting in greater impact per paper.

\subsection{Quality}

Paper awards are an indicator of
research quality. 
SOUPS has two award categories. One is given annually, including the IAPP SOUPS Privacy Award, Distinguished Paper Award, and Best Paper Award; the other given every several years, i.e. the SOUPS Impact Award.

\textbf{Annual award papers: }
There are 29 award papers between 2005 and 2024, with  
the following statistics:

Team Size: \textit{big} teams (23) significantly outnumbered \textit{small} teams (6), indicating that high-quality research tends to emerge from larger collaborative efforts.  

Author Gender: Male authors (81) accounted for 60\% of all authors, while female authors (54) made up 40\%.

Team Gender Composition: Mixed-gender teams (22) were the most common, followed by all-male teams (4) and all-female teams (3). 

First Author Gender: Male and female first authors were nearly equally represented (14 vs 15). 

Overall, these findings indicate that in first-category award-winning papers, larger team sizes and gender diversity are key characteristics of SOUPS high-quality research, highlighting the importance of collaboration and inclusivity in high-quality research.

\textbf{The SOUPS Impact Award: } 
At present, three papers have received this award (Table \ref{SOUPS Impact Award Papers} in the appendix), authored by a single-author team, a two-author team, and a six-author team. This contrasts with the trend of larger teams dominating annual paper awards. Further analysis of team gender composition shows that one paper was authored by a mixed-gender team, and two by all-male teams.

\section{Evolution of Community Structures}\label{sec:Evolution of Community Structures}

In this study, the scientific collaboration network of a specific conference is defined as a community, which consists of multiple connected components (referred to as islands). 
Each island is composed of nodes and edges, where nodes represent researchers (authors) and edges indicate collaboration relationships (co-authored papers) between researchers. Islands exhibit dynamic characteristics, evolving over time through growth (addition of new nodes or collaborations) or merging (fusion of multiple islands).

Community structure describes the distribution and organization of sub-networks within a scientific collaboration network, reflecting the overall characteristics of the network. 
By analyzing the number of collaborators per author, the growth trends in the number of islands, and changes in the sizes of the top three islands,  this section aims to reveal the evolutionary features of community structures in SOUPS and FC.

We analyzed the scientific collaboration networks in 2023 (Figure \ref{fig:Combined_Networks_2023} in the appendix).
In these networks, smaller islands at the periphery typically represent small research groups or individual collaborations, while larger islands at the center reflect broader collaboration networks. 
There are significant differences in network concentration and gender distribution between SOUPS and FC.
The largest island in SOUPS accounts for 51.73\% of all nodes, compared to 41.95\% in FC, indicating a more centralized network structure in SOUPS. Additionally, the figure clearly reveals differences in gender distribution: compared to SOUPS, FC has a significantly lower proportion of pink nodes, representing female authors.

\begin{figure}[h]
\centering
\includegraphics[width=\linewidth]{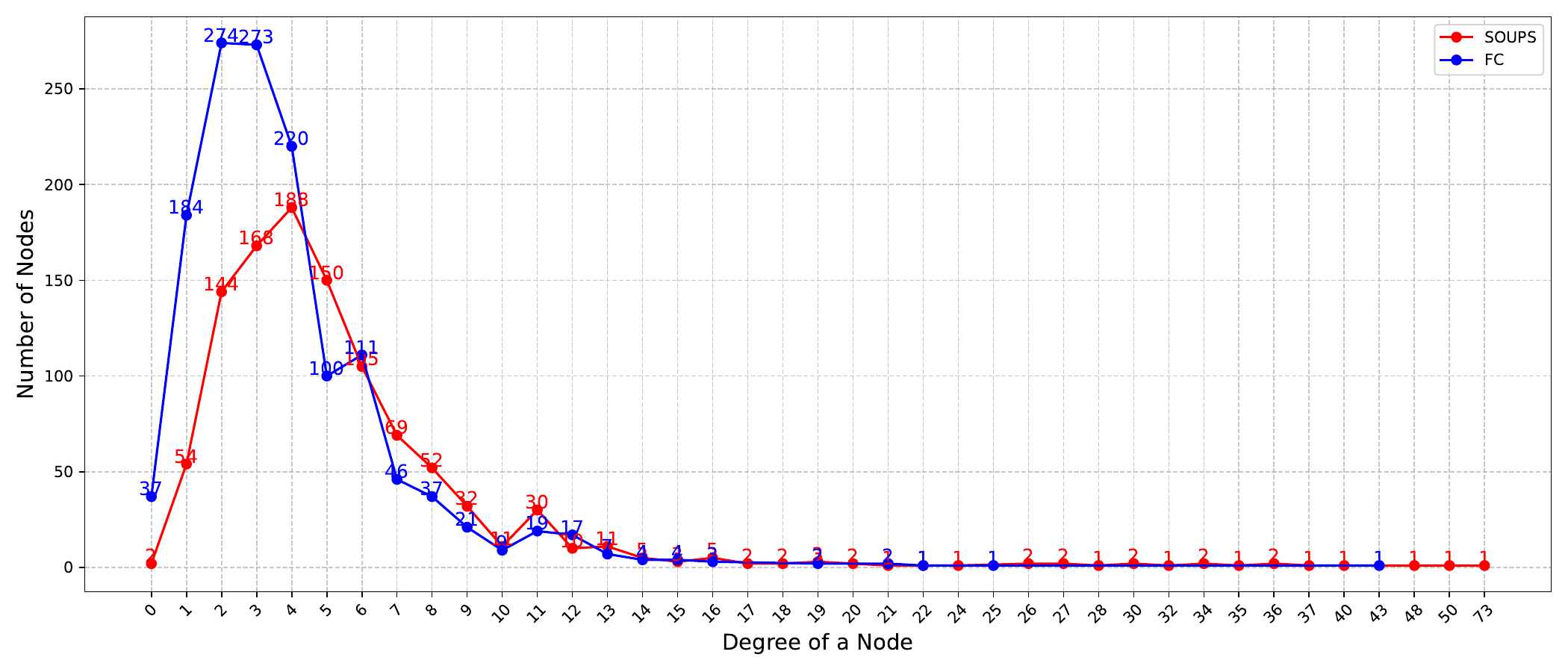}
\caption{The degree distribution. It shows the number of nodes and the degree of the node. 
}
\label{fig:Collaborator_Count_Distribution_SOUPS_FC}
\end{figure}

Regarding the distribution of author collaboration networks, Figure \ref{fig:Collaborator_Count_Distribution_SOUPS_FC} shows the number of collaborators per author in SOUPS and FC. The horizontal axis represents the degree of nodes in the collaboration network, with each node corresponding to an author, and the degree indicating the number of collaborators. The vertical axis represents the number of nodes, that is, the number of authors with the corresponding number of collaborators. 
Both conferences follow a similar trend: most authors have few collaborators, but a small proportion have many. 
Besides, Figure \ref{fig:Collaborator_Count_Distribution_SOUPS_FC} shows a difference in the most common number of collaborators between SOUPS and FC. In SOUPS, 4 collaborators are the most frequent, with 188 authors falling into this category, whereas in FC, 2 or 3 collaborators are the most common.

Authors with many collaborators play a crucial role in building broader collaboration networks. 
In SOUPS, the top five authors with the most collaborators are Lorrie Faith Cranor (73 collaborators), Alessandro Acquisti (50), Lujo Bauer (48), Michelle L. Mazurek (40), and Nicolas Christin (37). In FC, the top five authors are Ahmad-Reza Sadeghi (43 collaborators), Aggelos Kiayias (25), Pedro Moreno-Sanchez (22), Andrew Miller (21), and Moti Yung (21).

\begin{figure}[h]
\centering
\includegraphics[width=\linewidth]{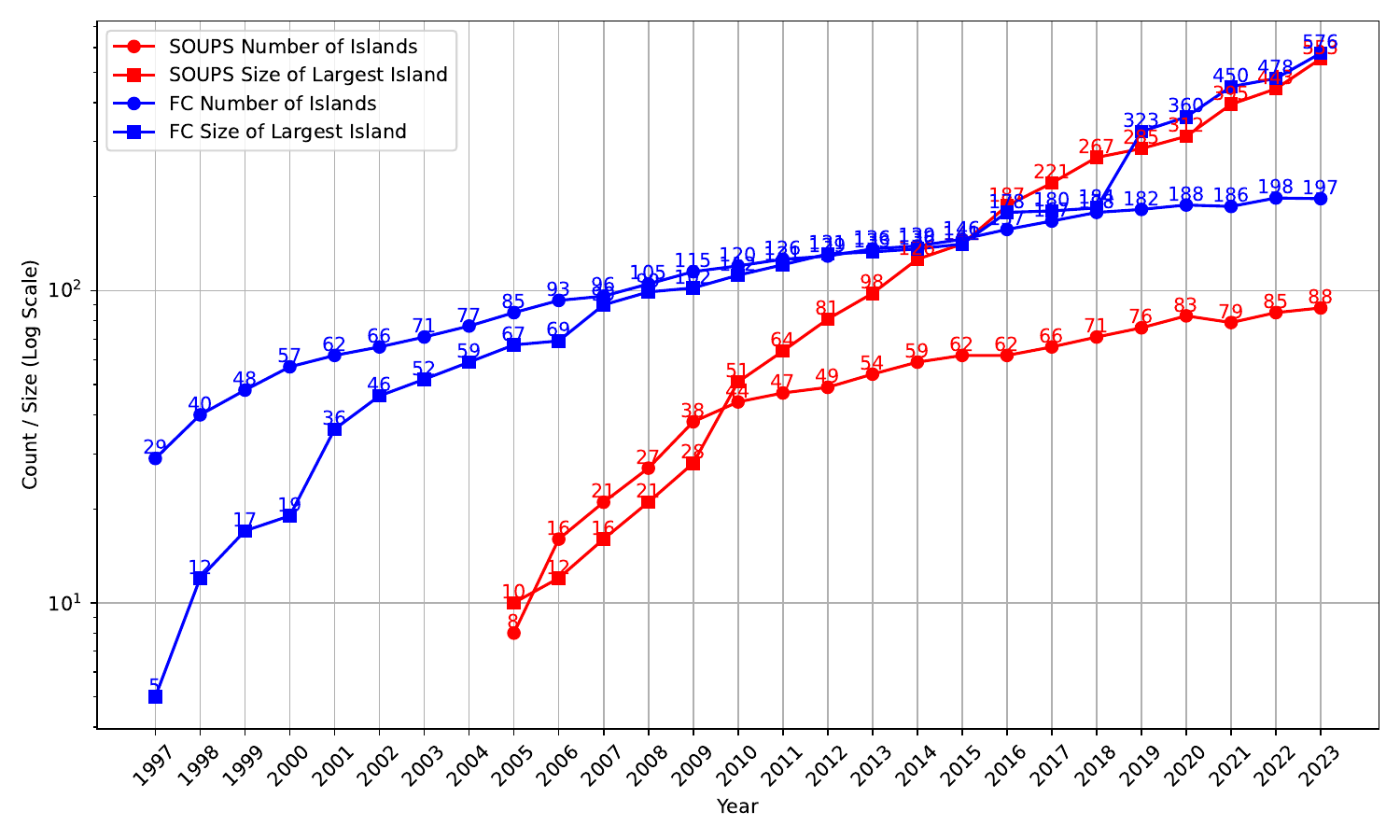}
\caption{Log-Scale analysis of collaboration networks: island counts and largest island sizes. 
}
\label{fig:Comparison_Log_Scale}
\end{figure}

Figure \ref{fig:Comparison_Log_Scale} illustrates the log-scale evolution of collaboration networks in SOUPS and FC, describing the changes in the number of islands and the size of the largest island over time. The vertical axis represents the total number of islands or the size of the largest island in terms of the number of nodes. Additionally, the vertical axis is presented on a logarithmic scale. 
From Figure \ref{fig:Comparison_Log_Scale}, it can be observed that the number of islands in both conferences shows an overall increasing trend, with FC exhibiting a faster growth rate. Regarding the size of the largest island, both conferences have experienced notable growth in recent years. In particular, FC had a sharp increase in its largest island's size in 2019. This was mainly due to the merger of two large islands; 
in the meanwhile, the largest island also grew itself by incorporating new authors.

\begin{figure*}[h]
\centering
\includegraphics[width=0.9\textwidth]{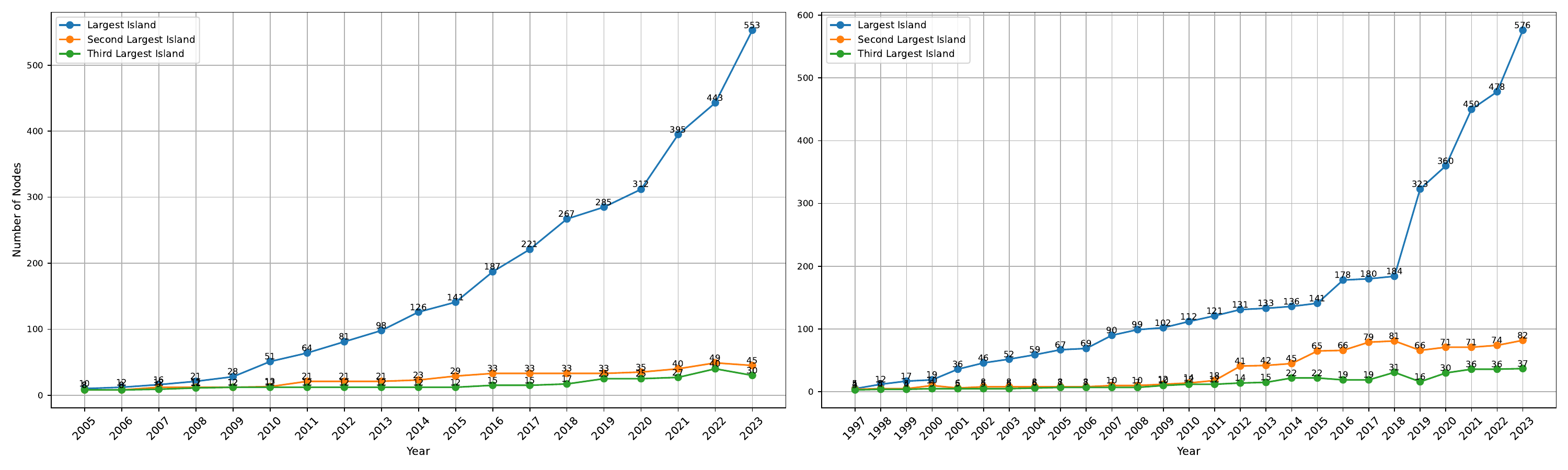}
\caption{The number of nodes in top 3 islands per year (Left: SOUPS; Right: FC). 
The largest island always grow; however the size of the 2nd or third largest islands may decrease, counterintuitively. 
}
\label{fig:Number_of_Nodes_in_Top_3_islands_per_Year_SOUPS_vs_FC}
\end{figure*}

Figure \ref{fig:Number_of_Nodes_in_Top_3_islands_per_Year_SOUPS_vs_FC} shows the changes in the number of nodes on the top three islands in SOUPS and FC over time.
In SOUPS, the size of the largest island shows steady and consistent growth over time.
In contrast, FC exhibits a notable shift in growth trends. Before 2018, the largest community in FC expanded at a moderate and steady pace. However, after 2018, the largest island experienced a rapid and significant expansion, primarily driven by the merging of smaller islands and the increasing convergence of researchers into a central network. 
This growth highlights a change in FC, with the central community becoming larger and more tightly connected.
Interestingly, the largest island always grow; however the size of the 2nd or third largest islands may decrease, counterintuitively. The explanation is simple: the previous 2nd or 3rd largest island was merged into the largest or 2nd largest island, and a new island emerged and took it over.

\section{Evolution of Research Topics}\label{sec:Evolution of Research Topics}

We adopted the method proposed by Milojevi~\cite{milojevic2015}, which identifies research topics by extracting and retaining key phrases from paper titles that reflect their core topics. For titles lacking clear thematic information, topics were manually assigned using domain expertise. Paper titles are highly concise and effectively reflect research content, making them an ideal source for identifying research topics. 
For simplicity, each topic is typically limited to no more than three words. Additionally, since some papers cover multiple themes, each paper can be associated with two or three topics.

Classic topic modeling like LDA~\cite{blei2003latent} requires extensive data  
but the Milojevic method uses only paper titles. 
We introduce several adjustments. For example, 
we merge some topics that are obviously the same but with different expressions.

\begin{figure}[h]
\centering
\includegraphics[width=\linewidth]{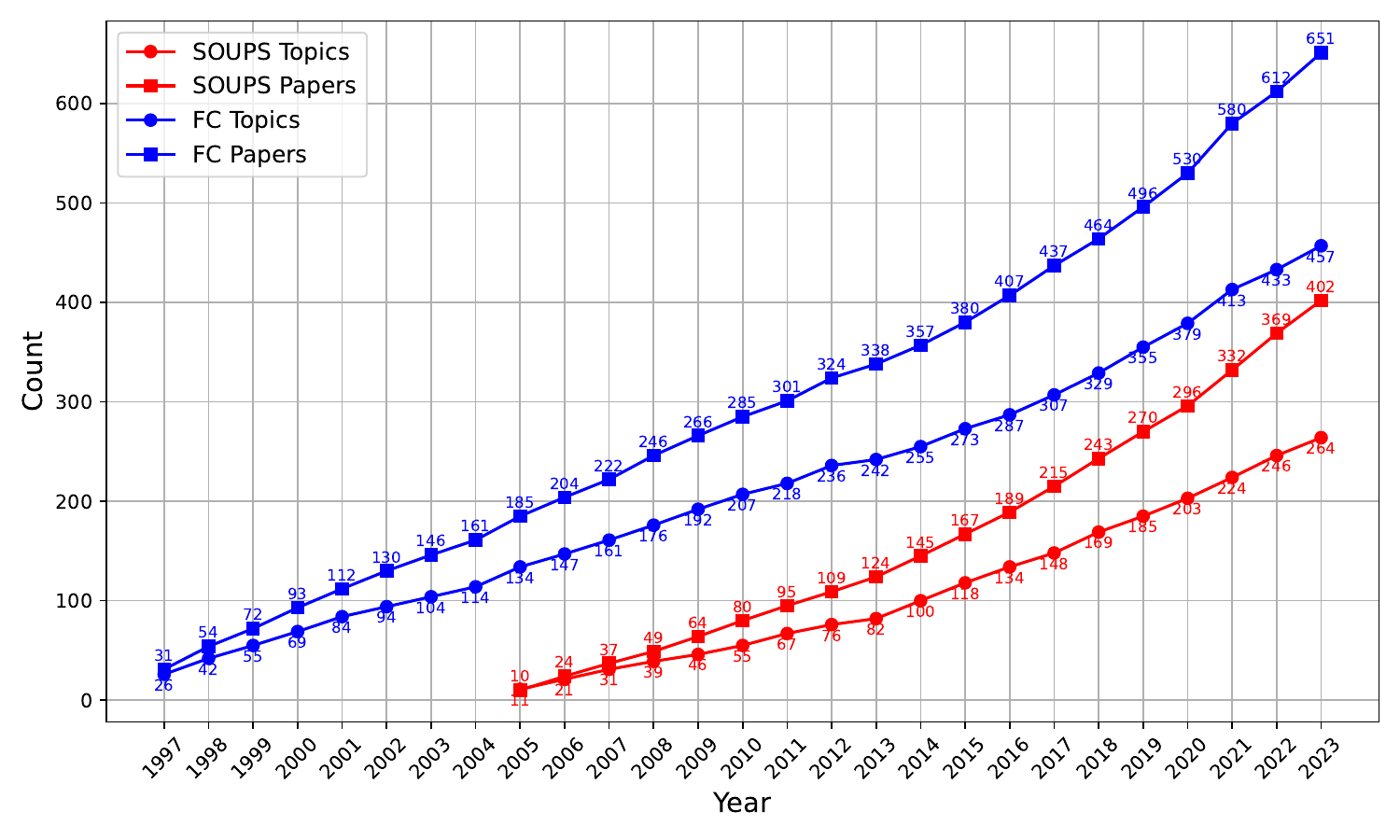}
\caption{Cumulative counts of research topics and papers 
}
\label{fig:Cumulative_Topics_and_Papers_SOUPS_vs_FC}
\end{figure}

Figure \ref{fig:Cumulative_Topics_and_Papers_SOUPS_vs_FC} illustrates the cumulative growth trends of the publication volume and unique research topics for SOUPS and FC.
First, the paper volume and the idea space appear to both increase exponentially for SOUPS and FC. Second, the paper volume grew much faster than the number of ideas in either SOUPS or FC.

We have generated word clouds (Figure \ref{fig:Word Cloud} in the appendix) for SOUPS and FC to clearly indicate which topics appear very frequently.
The topic phrases in SOUPS are predominantly focused on user authentication and privacy-related areas, with the top 15 most frequent phrases being authentication, password, phishing, graphical password, data sharing, access control, mental model, smart home, captcha, developer, device pairing, social network, software update, visual impairment, and data breach.
In contrast, The topic phrases in FC are primarily concentrated in the fields of financial cryptography and blockchain technologies. The top 15 most frequent phrases in FC are bitcoin, electronic cash, blockchain, digital signature, e-voting, authentication, anonymity, ethereum, micropayment, auction, market, secure multi-party computation, smart contract, privacy-preserving, and certificate revocation.

\section{Discussions}

\subsection{Structural patterns}

Our analysis revealed intriguing structural patterns, many of which are novel and some quite surprising. 

\textbf{Team Science.} 
Although the average number of authors per paper keeps increasing, two-author teams and three-author teams were most popular in FC, but three-author and four-author teams were most popular in SOUPS. The average number of authors per paper was constantly higher for SOUPS than FC every year.

We observe a positive correlation between the number of collaborators and the average number of publications for both SOUPS and FC. However, high productivity does not necessarily correlate with a high citation count for an author's most-cited paper.

\textbf{Gender dynamics.} Female authors have played a bigger role in SOUPS than in FC (Fig \ref{fig:Cumulative_Number_of_Male_and_Female_Authors_FC_vs_SOUPS} and \ref{fig:Male_and_Female_Authors_FC_vs_SOUPS_Percentage_with_Labels}). About 34.7\% of 1,069 SOUPS authors were female, whereas only 12.75\% of 1,373 FC authors were female. The cumulative growth of female authors was also much faster in SOUPS than in FC. 
There is a clear trend showing that the proportion of male authors in SOUPS has largely decreased over the years, while the proportion of female authors has steadily increased. By 2023, the proportion of female authors nearly equaled that of male authors 
(45.8\% vs 54.2\%).
In contrast, male authors have consistently dominated FC every year.

Gender composition in teams has manifested differently in SOUPS and FC. Its evolution has also differed, although we observe some similar trend. 
First, the proportion of all-male teams in SOUPS has largely followed a decreasing trend over the years, and the proportion of mixed-gender teams kept increasing. From 2005 to 2010, all-male teams and mixed-gender teams nearly took turn to dominate the SOUPS community. However, since 2011, mixed-gender teams have played an overwhelmingly dominating role in SOUPS.
Second, the proportion of all-male teams in FC has followed a decreasing trend in general, and the proportion of mixed-gender teams has followed an upward trend. However, all-male teams have been dominating in numbers over the years. 
We note that all-female teams have been rare in both SOUPS and FC. 

Males have overwhelmingly dominated the first-author role in FC every year, but SOUPS has shown a stark contrast. From 2005 to 2016, males dominated the first-author role in SOUPS most of the time. However, since 2017, the proportion of male and female first authors has been either equal or very close each year. 

\textbf{Team formation and performance.} We observe interesting correlations between team formation and citation counts. 

\textit{Team size.} 
The most popular team sizes in FC were 2 and 3, and these teams were also top performers in citation counts. The most popular team sizes in SOUPS were 3 and 4; however these teams were not top performers and they only performed relatively well. The top performers in SOUPS were teams of size 1 or 2, although their proportions were not high. 
We note that single-author teams were the worst performers in citation counts in FC.

\textit{Team size and gender composition.} 
The team formation with the highest average citations is
mixed-gender small teams for both SOUPS and FC. However, 
the most common are mixed-gender big teams for
SOUPS and all-male small teams for FC. Regardless of size, 
the mixed-gender teams achieved the highest average
citations in both conferences.

\textit{Team 
size and 1st author gender.} 
The most common are male first-author big teams in SOUPS and male
first-author small teams in FC. However, the team formation with the highest
average citations is female first-author small teams for both SOUPS and FC.

\textbf{Evolution of community structures.} In a collaborative network, many nodes eventually become connected, forming a so-called 'giant component' as defined by Newman~\cite{newman2001structure} (i.e. the largest island). SOUPS and FC are not an exception, which is unsurprising. 
However, we observed that the size of the largest island grew much faster in SOUSP than in FC. On the other hand, the number of islands in the FC community grew much faster than in SOUPS. We believe these are both a manifesto of the structural differences between SOUPS and FC. For example, 
the faster growth of the number of islands in FC is perhaps an echo of the fact that smaller teams have constantly done well in FC, whereas larger teams were more popular in SOUPS. 

It is well known that in the evolution of a collaborative network, the majority of nodes have only a small number of collaborators, and a minority will have developed a large number of collaborators. This pattern is mathematically characterized by a power-law distribution (or a long-tail distribution)~\cite{newman2001structure}. Both SOUPS and FC exhibited this pattern to a large extent. 

\textbf{Evolution of research themes.}
It is 
conventional wisdom that the number of papers increased exponentially (with an average doubling period of 15 years), but the number of ideas increased only linearly with time~\cite{fortunato2018science, milojevic2015}. 
We did not observe this phenomenon, and our observations are the following, instead. First, the paper volume and the idea space appear to both increase exponentially for SOUPS and FC. We believe that this can be explained by the fact that both SOUPS and FC represented emerging and fast-moving disciplines, and their growth was at an extraordinary rate given their young ages. Still, the paper volume grew much faster than the number of ideas in either SOUPS or FC.

\textbf{Why different?} Many of the differences may be explained by the differences in research problems 
and topics in these two communities. Consequently they require different research methods, skillsets and suitable team formations.

\subsection{Implication and Impact}

Our analysis can inform multiple stakeholders such as faculty members, team leaders, future thought leaders, department and institute heads, and even funding agencies. 

\textbf{Team building. }Our results readily suggest faculty members or team leaders some rational and strategic considerations for their team building. For example, if you aim for SOUPS, you should either have a relatively large lab or actively collaborate with other teams. The average number of authors per papers is a reasonable estimate of the amount of work expected for a single paper. This average number keeps increasing with time and it is also consistently larger in SOUPS than in FC. Moreover, the proportion of teams of four or more authors in SOUPS have reached nearly 80\%. In general, it is hard
for a small lab to compete with large labs in terms of both productivity and impact, and it may get much harder in this community
in the future.

If you aim for FC or similar communities, the head count of your lab is less critical than SOUPS. Although the proportion of larger teams has increased in FC in the recent years, smaller teams have performed well consistently in this community over the years. We are not saying that the advantage of large labs does not apply to FC.

We note that two-author teams have been among top performers in average paper citations in both SOUPS and FC. This suggests that a viable alternative approach is to run small but good teams, and write less but better papers. This way, a faculty member will have more time to work as a scientist than as a manager. 

In terms of citation counts, mixed gender small teams have been the top performers in both SOUPS and FC. Therefore, it looks like a reasonable strategy to aim for a mixed-gender lab for both SOUPS and FC. However, we reckon that recruiting female researchers in FC is more challenging than in SOUPS, given its consistently low ratio of females.

\textbf{Faculty recruit.} Cybersecurity is usually in the camp of STEMs, where females (including students and faculty members) are less represented than their male counterparts. Our results suggest that FC is inline with this stereotype, but SOUPS appears to be an outlier. This has policy implications such as in faculty recruitment. Hiring a new faculty member in usable security is more likely than one in FC to increase the gender balance in a department or institute, even if the new hire is male himself. The reasoning is simple. To thrive in the SOUPS community, his team will more likely hire female members (including both postdocs and PhD students) than he would work in FC or other security sub-disciplines. 
The ratio of female authors, female first authors, mixed-gender high-performing teams was all much higher in SOUPS than that in FC. Therefore, SOUPS is likely an environment more conducive than FC for female researchers to grow themselves and thrive. 
This will not only increase gender diversity in his own team, but  contribute to diversity of his department and institute in the long run.

\textbf{Funding agencies. }For funding agencies interested in fostering an inclusive research community, our research shows that gender disparity does not manifest uniformly across the board. Instead, some security communities may do better than the others. Therefore, as an example, if the agencies 
aim to support academic conferences to widen participation by female students and researchers, our study gives some hints where they will be able to find good candidates, and where more efforts will be required to encourage gender parity.

\textbf{Thought leaders. }For future thought leaders, perhaps the most important lesson is that our method reveals informative and sometimes hidden patterns, structures and dynamics, all of which could help understand deeply a research community and its growth. These understandings may guide and predicate the evolution of a new research field. However, such an analysis is unlikely to be possible until sufficient meta-data become available.

\subsection{The Gender Prediction algorithm}

The CCT algorithm~\cite{van2023open} was well conceived, with an elegant mathematical structure. It has a good extensibility by integrating new datasets. 

Our simple improvement in preprocessing enhances both accuracy and coverage of the classifier. However, our improvement only addresses the weakness of the software, and it does not pinpoint any fundamental issues. 

Some subtle issues have to be considered in applying the CCT. For example, the trade-off between predication and accuracy. Second, 
the most likely prediction by Bayes' and the consensus is mathematically sound, but it may not agree with the reality of a particular dataset. As an illustrative example, unisex name Moni as in Hebrew is more likely predicted by the CCT as female, however, Moni Naor will for sure be a male name if it occurs in a dataset of security people.

\textbf{Further improvements. } 
The CCT classifier has already 
incorporated useful information such as country, culture and era to improve its accuracy. 
We observe that in certain disciplines with significant gender disparities, the distribution of genders may exhibit a clear bias. For example, if the majority of authors in a specific discipline are male, even if the CCT classifier predicts certain names as neutral, incorporating discipline information could reasonably infer that the author is more likely male. This suggests that discipline can serve as an auxiliary variable to enhance the accuracy of gender predictions.

Conventional perspectives assume that information such as discipline is unrelated to gender and thus not used for gender prediction. However, in reality, these contextual factors (e.g., discipline) do contain and be able to convey gender-related clues. We believe this is a valuable future direction for further improving the CCT.

We note that the CCT algorithm is limited to binary gender classification only.

\section{Conclusion}

Via an analysis of meta data of published conferences papers only, we uncovered novel and some surprising structures in the communities of SOUPS and FC. Perhaps the most interesting results are the vast differences in gender dynamics in these communities, and the team formation
with the highest average citations is mixed-gender small teams
for both SOUPS and FC. We have discussed the implication and impact of our findings on actionable practices and policy.

We also independently validated a state-of-the-art algorithm for predicting gender only by name, improved it and proposed further enhancements. All these have general value, in particular for future science of science studies. 

Our study can be adopted to understand the evolution of other cybersecurity subfields. It can also be applied to study the four premier conferences in security such as CCS, NDSS, USENIX Security, and IEEE S\&P. Since these conferences have a long history, cover a broad range of research topics and feature most if not all major players, a meta analysis like ours will likely produce an interesting and complete picture of the evolution of cybersecurity research as a dynamic and fast-moving discipline.

\newpage

\section*{Ethics considerations}
This study is a meta-science research project that examines the practice of science itself. It does not involve any ethical risks or direct participation of individuals. The study relies entirely on secondary data, with all collected information obtained through lawful means from public repositories. We consulted our university Research Ethics Committee, which reviewed and approved that the study complies fully with ethical guidelines. By adhering strictly to ethical standards and best practices, this research ensures transparency, legality, and integrity throughout its process.

\section*{Open science}
Open Science is an initiative aimed at sharing research findings, data, and tools in an open and transparent way to promote scientific progress. We fully support this initiative and have integrated its principles into our research practices. Therefore, we commit to following Open Science policies. Once the paper is accepted, we will promptly upload the research data to an open repository with detailed metadata and documentation. Additionally, we will open-source the code used in the research and host it on GitHub or other open platforms to enhance transparency and reproducibility.

\bibliographystyle{plain}
\bibliography{references}

@misc{dblp,
  title = {{DBLP}},
  howpublished = {\url{https://dblp.dagstuhl.de/}}
}

@misc{usenix,
  title = {{USENIX}},
  howpublished = {\url{https://www.usenix.org/}}
}

@article{uzzi2013atypical,
  title={Atypical combinations and scientific impact},
  author={Uzzi, Brian and Mukherjee, Satyam and Stringer, Michael and Jones, Ben},
  journal={Science},
  volume={342},
  number={6157},
  pages={468--472},
  year={2013},
  publisher={American Association for the Advancement of Science}
}

@article{sinatra2016quantifying,
  title={Quantifying the evolution of individual scientific impact},
  author={Sinatra, Roberta and Wang, Dashun and Deville, Pierre and Song, Chaoming and Barab{\'a}si, Albert-L{\'a}szl{\'o}},
  journal={Science},
  volume={354},
  number={6312},
  pages={aaf5239},
  year={2016},
  publisher={American Association for the Advancement of Science}
}

@article{milojevic2015,
  title={Quantifying the cognitive extent of science},
  author={Milojevi{\'c}, Sta{\v{s}}a},
  journal={Journal of Informetrics},
  volume={9},
  number={4},
  pages={962--973},
  year={2015},
  publisher={Elsevier}
}

@article{wuchty2007increasing,
  title={The increasing dominance of teams in production of knowledge},
  author={Wuchty, Stefan and Jones, Benjamin F and Uzzi, Brian},
  journal={Science},
  volume={316},
  number={5827},
  pages={1036--1039},
  year={2007},
  publisher={American Association for the Advancement of Science}
}

@article{wu2019large,
  title={Large teams develop and small teams disrupt science and technology},
  author={Wu, Lingfei and Wang, Dashun and Evans, James A},
  journal={Nature},
  volume={566},
  number={7744},
  pages={378--382},
  year={2019},
  publisher={Nature Publishing Group UK London}
}

@article{bromham2016interdisciplinary,
  title={Interdisciplinary research has consistently lower funding success},
  author={Bromham, Lindell and Dinnage, Russell and Hua, Xia},
  journal={Nature},
  volume={534},
  number={7609},
  pages={684--687},
  year={2016},
  publisher={Nature Publishing Group UK London}
}

@inproceedings{van2023open,
  title={An open-source cultural consensus approach to name-based gender classification},
  author={Van Buskirk, Ian and Clauset, Aaron and Larremore, Daniel B},
  booktitle={Proceedings of the International AAAI Conference on Web and Social Media},
  volume={17},
  pages={866--877},
  year={2023}
}

@article{blei2003latent,
  title={Latent dirichlet allocation},
  author={Blei, David M and Ng, Andrew Y and Jordan, Michael I},
  journal={Journal of machine Learning research},
  volume={3},
  number={Jan},
  pages={993--1022},
  year={2003}
}

@article{fortunato2018science,
  title={Science of science},
  author={Fortunato, Santo and Bergstrom, Carl T and B{\"o}rner, Katy and Evans, James A and Helbing, Dirk and Milojevi{\'c}, Sta{\v{s}}a and Petersen, Alexander M and Radicchi, Filippo and Sinatra, Roberta and Uzzi, Brian and others},
  journal={Science},
  volume={359},
  number={6379},
  pages={eaao0185},
  year={2018},
  publisher={American Association for the Advancement of Science}
}

@article{newman2001structure,
  title={The structure of scientific collaboration networks},
  author={Newman, Mark EJ},
  journal={Proceedings of the national academy of sciences},
  volume={98},
  number={2},
  pages={404--409},
  year={2001},
  publisher={National Acad Sciences}
}
\section*{Appendix}

\begin{figure*}[h]
\centering
\begin{subfigure}[b]{0.45\textwidth} 
    \centering
    \includegraphics[width=\linewidth]{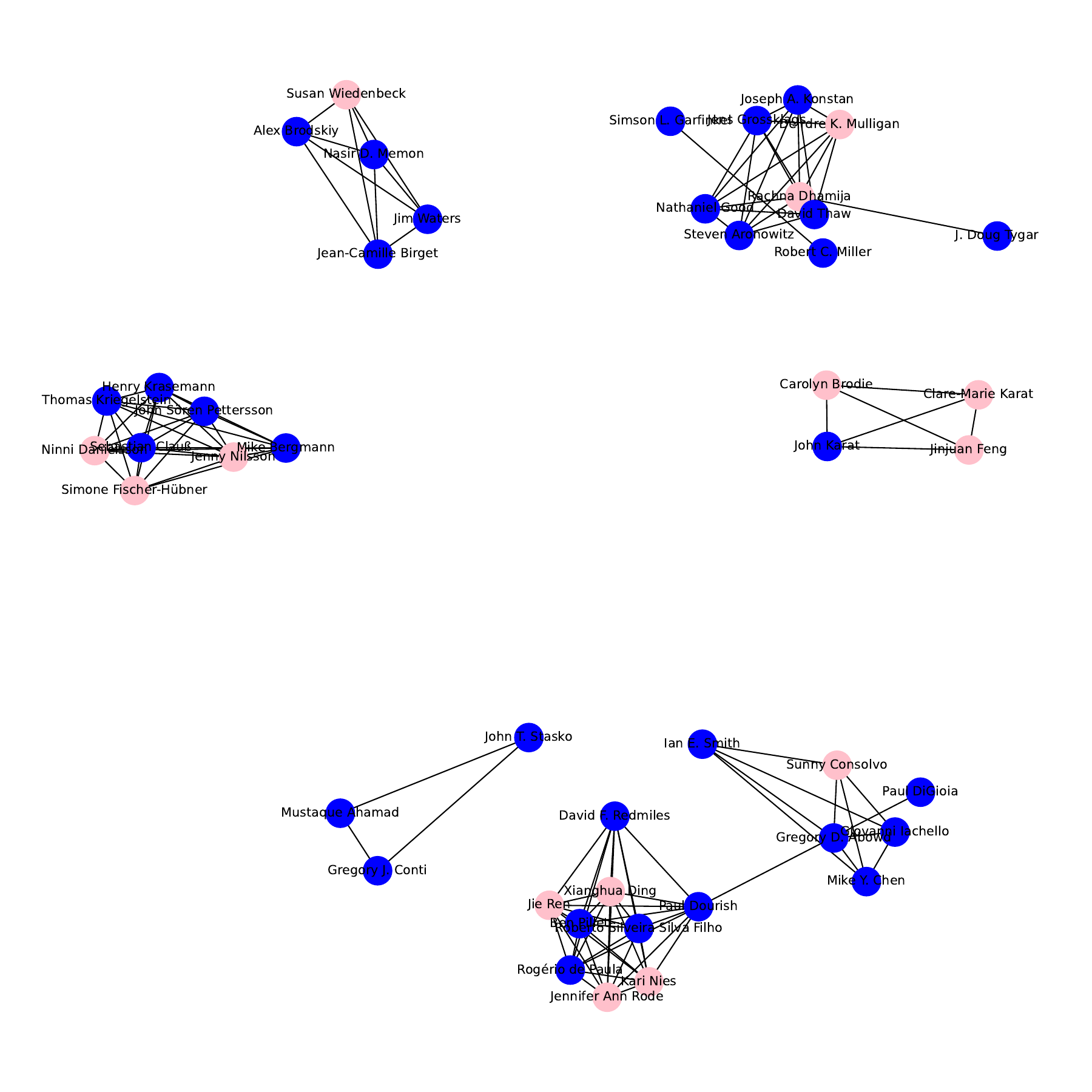}
    \label{fig:Screenshot_SOUPS_2005}
\end{subfigure}
\hspace{0.02\textwidth} 
\begin{subfigure}[b]{0.45\textwidth}
    \centering
    \includegraphics[width=\linewidth]{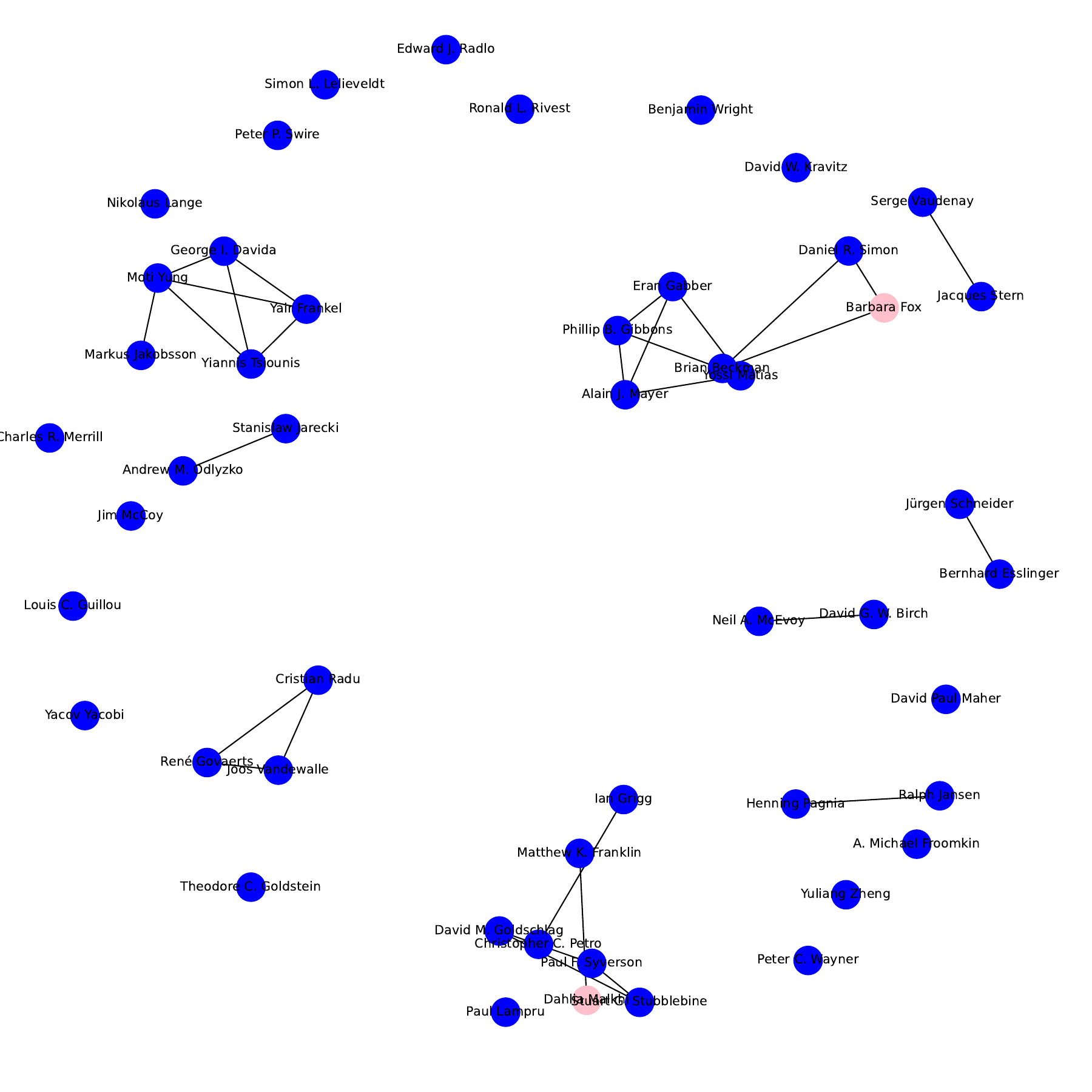}
    \label{fig:Screenshot_FC_1997}
\end{subfigure}
\caption{Scientific collaboration networks in the first year (Left: SOUPS; Right: FC). In comparison with Figure \ref{fig:Combined_Networks_2023}, these visualizations highlight the significant evolution of scientific collaboration networks over the years.}
\label{fig:Combined_Networks_begin}
\end{figure*}

\begin{figure*}[h]
\centering
\begin{subfigure}[b]{0.45\textwidth} 
    \centering
    \includegraphics[width=\linewidth]{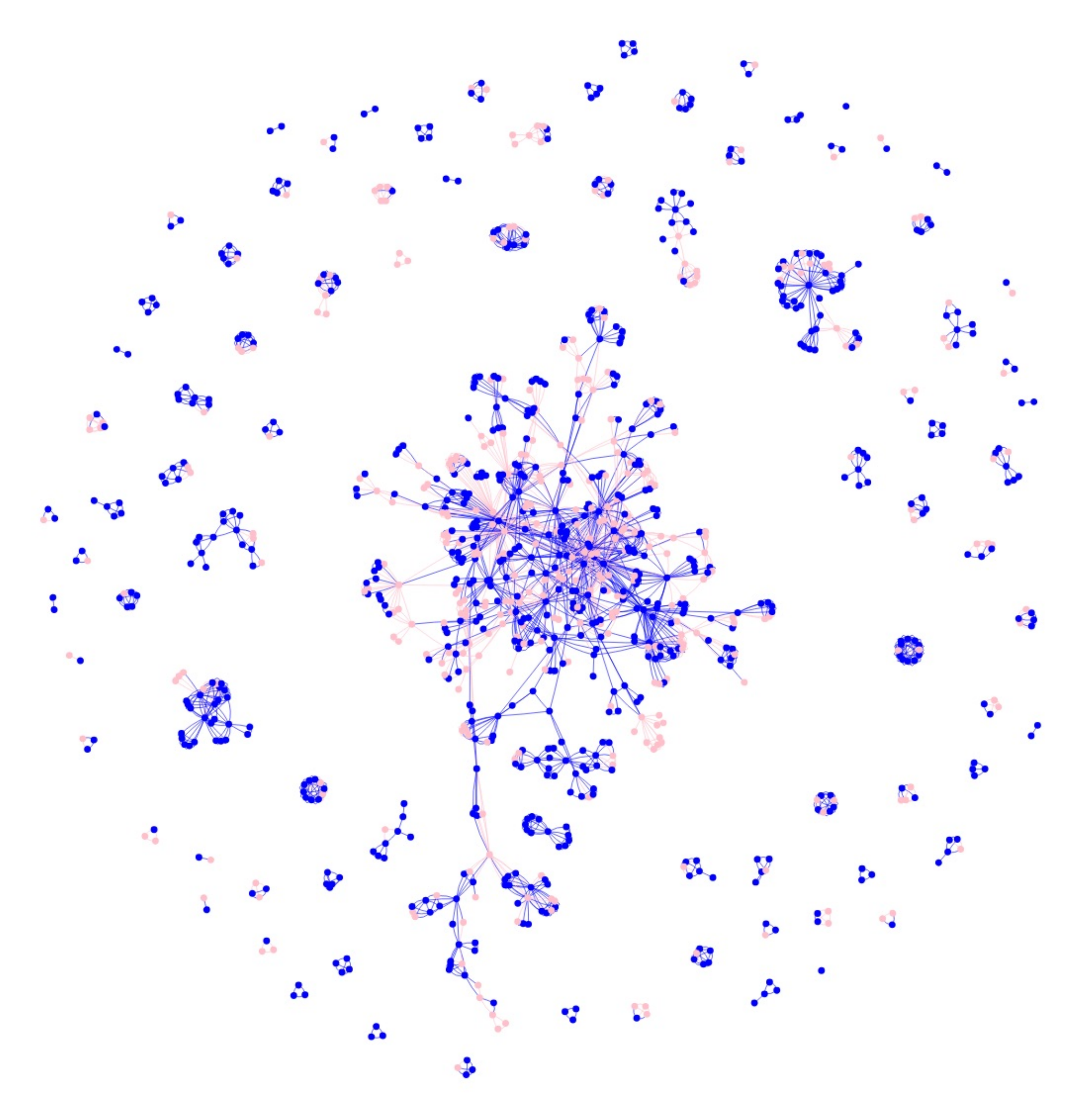}
    \label{fig:Screenshot_SOUPS}
\end{subfigure}
\hspace{0.02\textwidth} 
\begin{subfigure}[b]{0.45\textwidth}
    \centering
    \includegraphics[width=\linewidth]{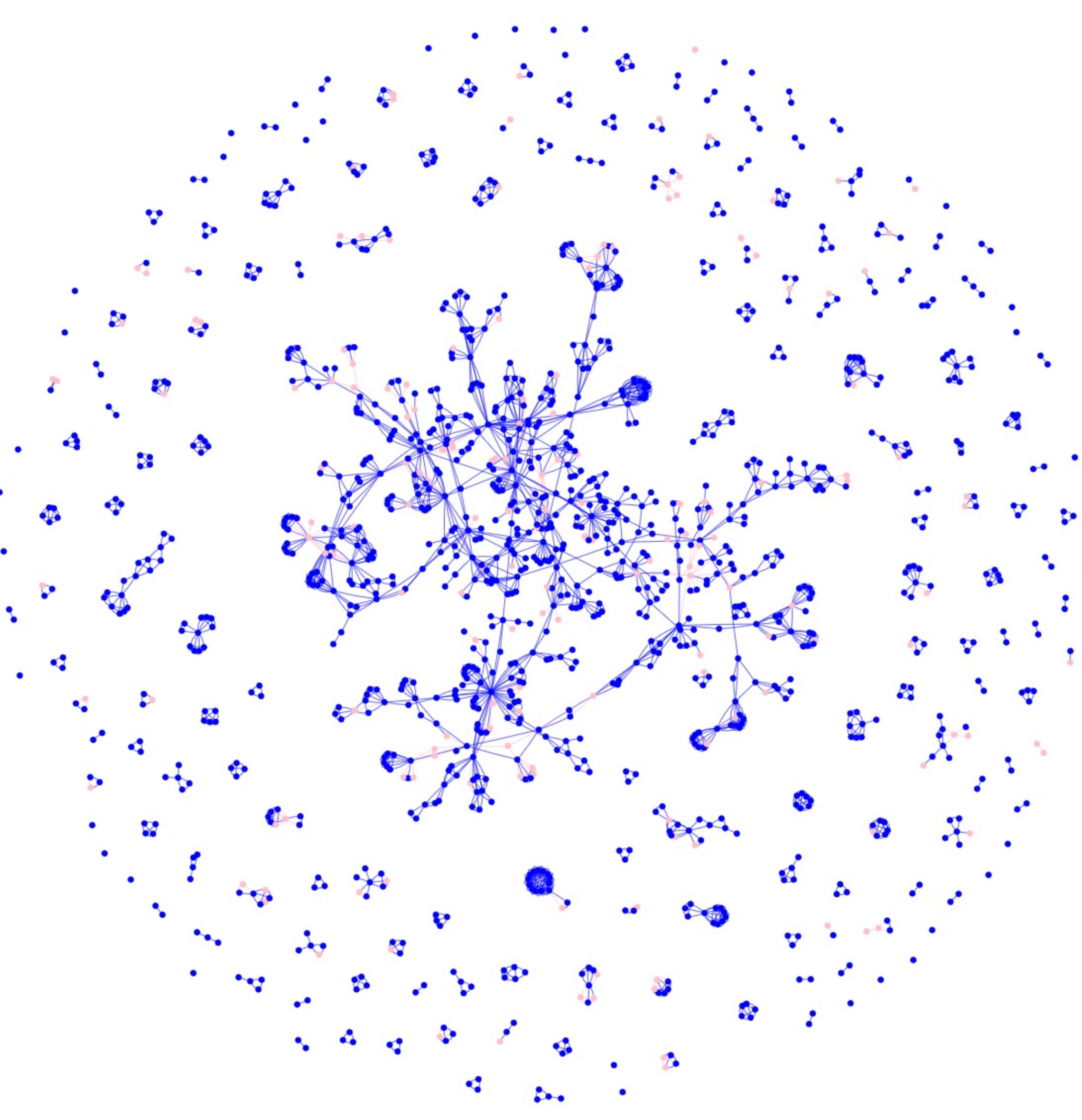}
    \label{fig:Screenshot_FC}
\end{subfigure}
\caption{Collaboration networks in 2023 (Left: SOUPS; Right: FC). Edges represent collaboration relationships between researchers. Pink nodes represent female authors, while blue nodes represent male authors.}
\label{fig:Combined_Networks_2023}
\end{figure*}

\begin{figure*}[h]
\centering
\begin{subfigure}[b]{0.45\textwidth}
    \centering
    \includegraphics[width=\linewidth]{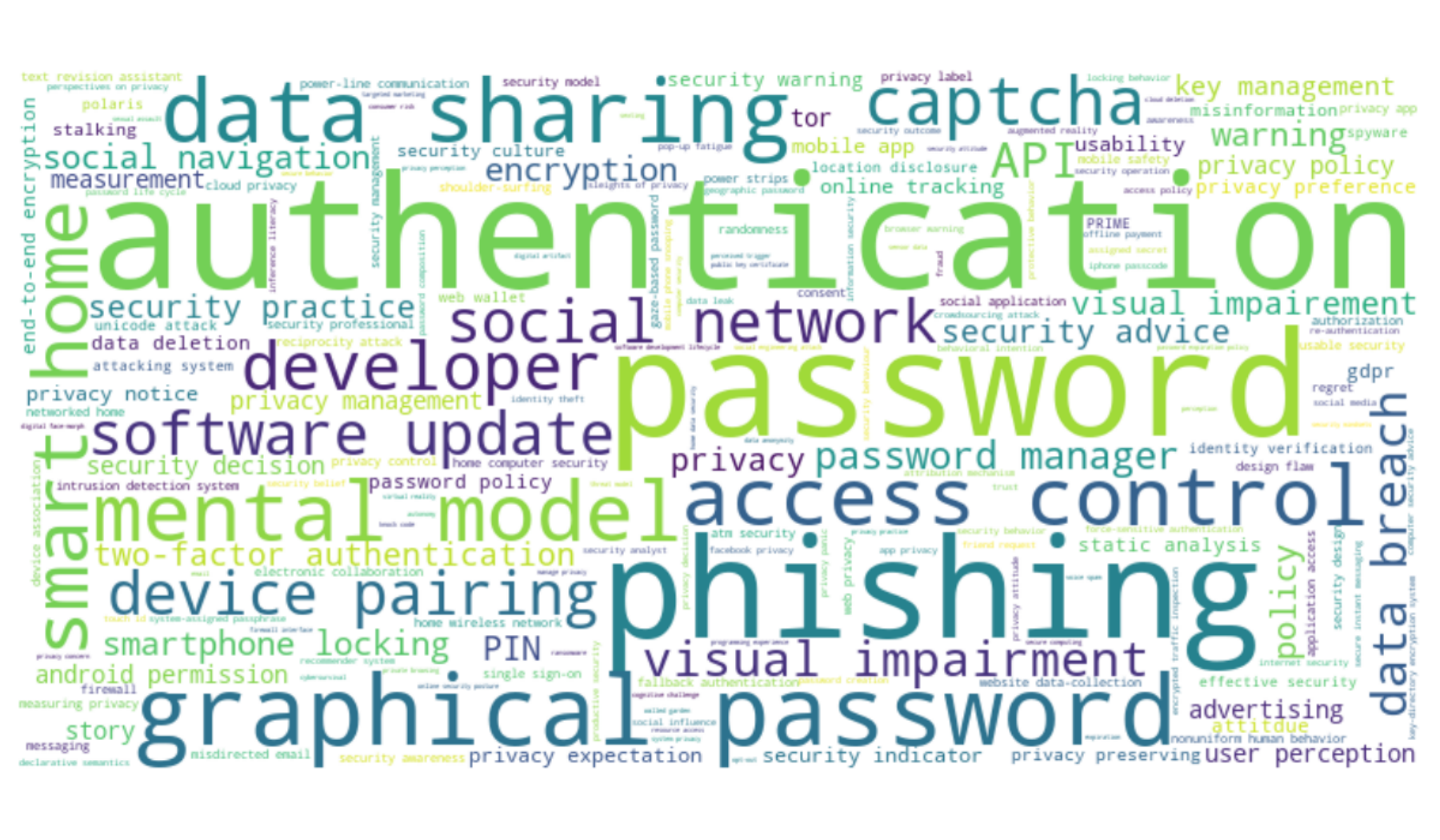}
\label{fig:SOUPS_Word Cloud by Topic Phrases}
\end{subfigure}
\hspace{0.02\textwidth} 
\begin{subfigure}[b]{0.45\textwidth}
    \centering
    \includegraphics[width=\linewidth]{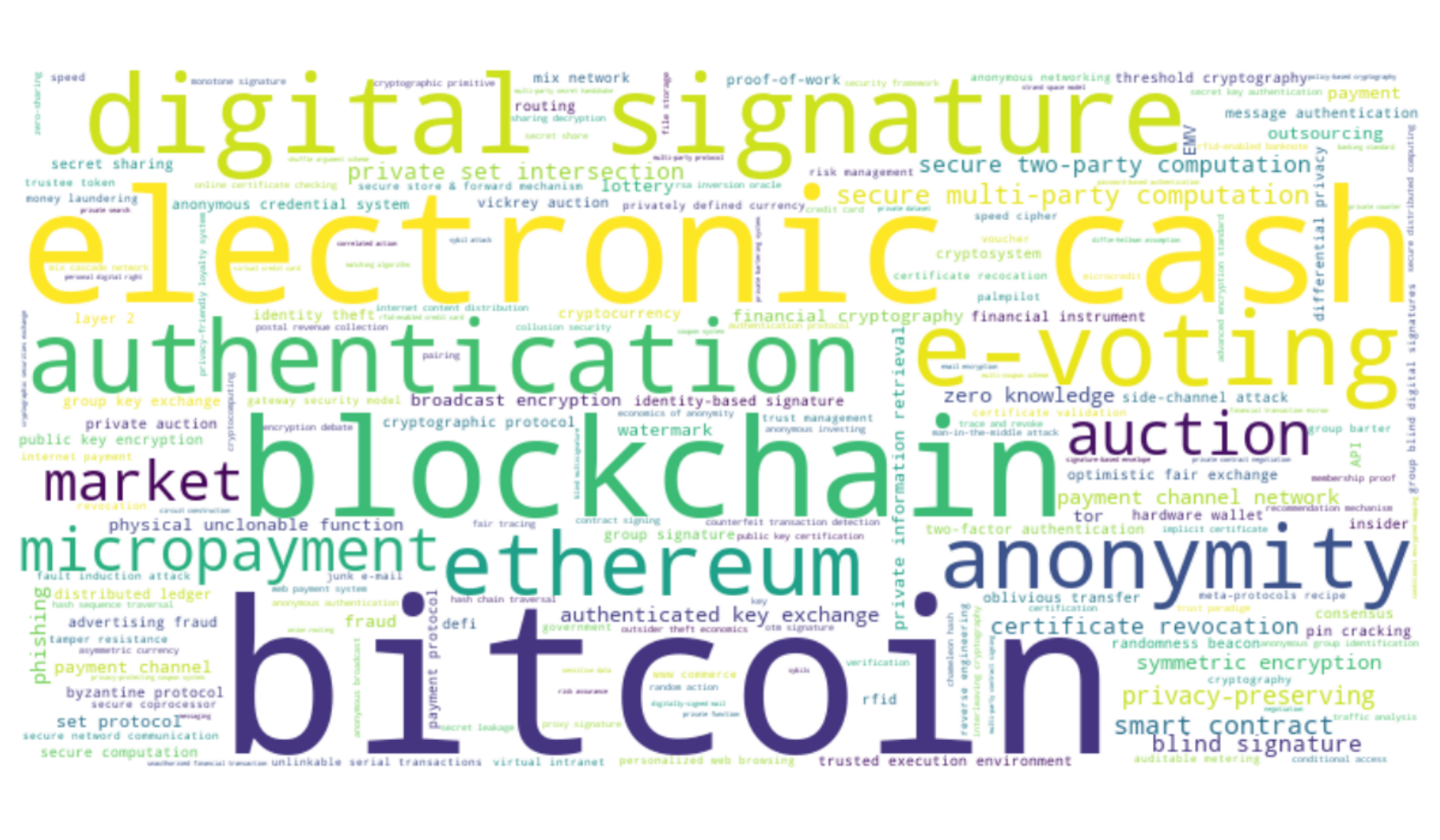}
    \label{fig:FC_Word Cloud by Topic Phrases}
\end{subfigure}
\caption{Word cloud by topic phrases (Left: SOUPS; Right: FC). The larger keywords indicate higher frequencies of occurrence, reflecting the dominant topics discussed in each community. }
\label{fig:Word Cloud}
\end{figure*}

\begin{figure*}[h]
    \raggedright
    \rotatebox{90}{
        \begin{minipage}{\textheight} 
            \centering            \includegraphics[width=\textwidth]{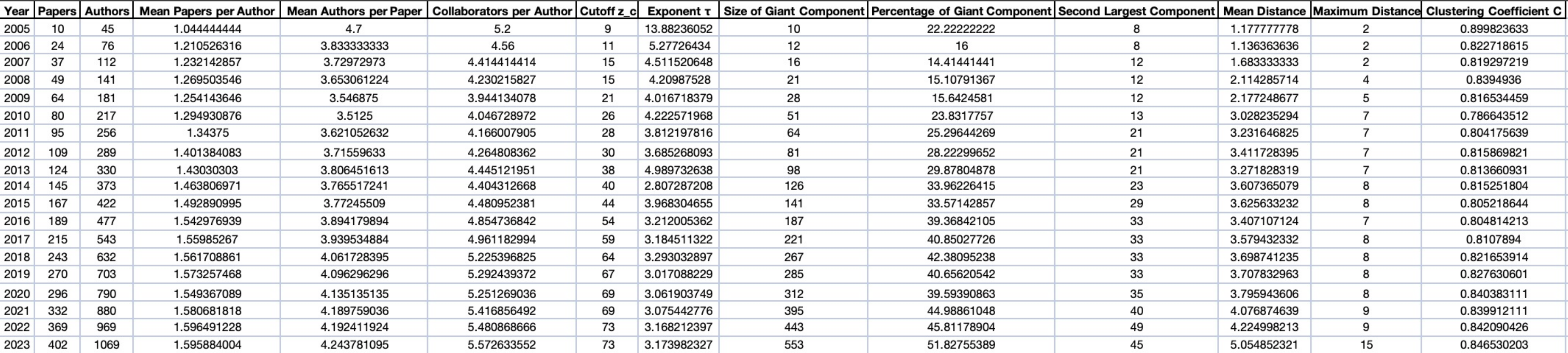} 
            \caption{SOUPS indicators.  The growth in the number of papers and authors highlights the increasing influence and attention that the SOUPS conference commands within the academic community. The average number of authors per paper remains stable at around 4, indicating that team collaboration is the predominant model, with typical team sizes consisting of about 4 authors. The increase in $z\_c$ and the decrease in $\tau$ suggest that supernodes play an increasingly significant role in the network. The expansion of the giant component and the high clustering coefficient indicate that the overall connectivity and cohesiveness of the collaboration network are improving. Meanwhile, the average path length and clustering coefficient reflect the network’s local cohesiveness and “small-world” characteristics. These metrics collectively provide a more comprehensive understanding of the dynamic changes and structural features of the SOUPS scientific collaboration network.
    }
            \label{fig:SOUPS indicators}
        \end{minipage}
    }
\end{figure*}

Table \ref{SOUPS Impact Award Papers} lists the SOUPS Impact Award papers.
Figure \ref{fig:Combined_Networks_begin} illustrates the scientific collaboration networks in SOUPS during 2005 and FC during 1997. 
Figure \ref{fig:Combined_Networks_2023} shows the scientific collaboration network between SOUPS and FC in 2023.
Figure \ref{fig:Word Cloud} shows word clouds generated using topic phrases to clearly indicate which topics are high-frequency topics.
Figure \ref{fig:SOUPS indicators} draws on the research methods and metrics used in a classic paper by Newman to analyze the structure and evolution of the scientific collaboration network within the SOUPS conference~\cite{newman2001structure}. This table summarizes key statistical data and network characteristics of the SOUPS collaboration network over multiple years, including the total number of papers and authors, the average number of papers per author, the average number of authors per paper, the average number of collaborators per author, the cutoff point and exponent of the power-law distribution, the size and proportion of the giant component, the size of the second-largest component, the average shortest path length, the maximum path length between nodes in the network, and the clustering coefficient.

\begin{table}[h]
    \centering
    \renewcommand{\arraystretch}{1.5}
    \small
    \begin{tabular}{p{0.3cm}p{2.9cm}p{2.5cm}p{1.4cm}}
        \toprule
        \textbf{Year} & \textbf{Title} & \textbf{Name} & \textbf{Gender} \\ 
        \midrule
        2020 & Folk Models of Home Computer Security & Rick Wash & Male \\ 
        2017 & Android Permissions: User Attention, Comprehension, and Behavior & Adrienne Porter Felt, Elizabeth Ha, Serge Egelman, Ariel Haney, Erika Chin, David A. Wagner & Female, Female, Male, Female, Female, Male \\ 
        2014 & Usability of CAPTCHAS Or usability issues in CAPTCHA design & Jeff Yan, Ahmad El Ahmad & Male, Male \\ 
        \bottomrule
    \end{tabular}
    \caption{SOUPS Impact Award papers.}
    \label{SOUPS Impact Award Papers}
\end{table}

\end{document}